\documentclass[11pt]{article}

\usepackage[margin=2.5cm]{geometry}
\usepackage{amsmath}
\usepackage{graphicx}
\usepackage{caption}
\usepackage{setspace}
\usepackage{multirow}
\usepackage{tabularx}
\usepackage{bbold}
\usepackage{indentfirst}
\usepackage{multirow}
   
\usepackage{fullpage,times,subfigure,fancyhdr}

\usepackage{sidecap}
\usepackage{wrapfig}

\usepackage[colorlinks=true,pagebackref,linkcolor=magenta]{hyperref}

\usepackage[margin=2.5cm]{geometry}
\usepackage{amsmath}
\usepackage{graphicx}
\usepackage{caption}
\usepackage{setspace}
\usepackage{multirow}
\usepackage{tabularx}
\usepackage{bbold}
\usepackage{indentfirst}
\usepackage{natbib} 
\usepackage{soul, color}
\pdfoutput=1
\usepackage{sidecap}
\usepackage{wrapfig}

\doublespace

\title{Relating multi-sequence longitudinal intensity profiles and clinical covariates in new multiple sclerosis lesions}
\author{Elizabeth M. Sweeney\textsuperscript{1,2}, Russell T. Shinohara\textsuperscript{3},  Blake E. Dewey\textsuperscript{2}, Matthew K. Schindler\textsuperscript{2}, \\ John Muschelli\textsuperscript{1}, Daniel S. Reich\textsuperscript{1,2,4,5}, Ciprian M. Crainiceanu\textsuperscript{1},  Ani Eloyan\textsuperscript{1}
\\ }

\begin{document}

\maketitle

\small \noindent \textsuperscript{1}Department of Biostatistics, The Johns Hopkins University, Baltimore, MD 21205

\noindent \textsuperscript{2}Translational Neuroradiology Unit, Division of Neuroimmunology and Neurovirology, National Institute of Neurological Disease and Stroke, National Institute of Health, Bethesda, MD 20892 

 \noindent \textsuperscript{3}Department of Biostatistics and Epidemiology, Center for Clinical Epidemiology and Biostatistics, Perelman School of Medicine, University of Pennsylvania, Philadelphia, PA 19104

\noindent \textsuperscript {4}Department of Radiology, The Johns Hopkins University School of Medicine, Baltimore, MD 21287

\noindent \textsuperscript {5}Department of Neurology, The Johns Hopkins University School of Medicine, Baltimore, MD 21287

\bigskip

\bigskip

\bigskip

\noindent Corresponding Author: 

\noindent Elizabeth Sweeney 

\noindent (317)698-5700

\noindent emsweene@jhsph.edu
 
\normalsize
\bigskip

\newpage

\section*{Abstract} 
Structural magnetic resonance imaging (MRI) can be used to detect lesions in the brains of multiple sclerosis (MS) patients. The formation of these lesions is a complex process involving inflammation, tissue damage, and tissue repair, all of which are visible on MRI and potentially modifiable by pharmacological therapy, including neuroprotective and reparative agents now in development.   Here we characterize the lesion formation process on longitudinal, multi-sequence structural MRI from 34 MS patients, scanned on average once every 37 days (sd 52.3, range [13, 889]), with an average  of 21 scans each (sd 8.0, range [10, 37]).  We then relate the longitudinal changes we observe within lesions to disease subtype and therapeutic interventions.  In this article, we first outline a pipeline to extract voxel level, multi-sequence longitudinal profiles from four MRI sequences within lesion tissue.  We then propose two models to relate clinical covariates to the longitudinal profiles.   The first model is a principal component analysis (PCA) regression model, which collapses the information from all four profiles into a single scalar value.  We find that the score on the first PC identifies areas of slow, long-term intensity changes within the lesion at a voxel level, as validated by two experienced clinicians (a neuroradiologist and a neurologist).  On a quality scale of 1 to 4, with 4 being the highest, the neuroradiologist gave the score on the first PC a median rating of 4 (95\% CI: [4,4]), and the neurologist gave it a median rating of 3 (95\% CI: [3,3]).  In the PCA regression model, we find that treatment with disease modifying therapies (p-value $<$ 0.01), steroids (p-value $<$ 0.01), and being closer to the boundary of abnormal signal intensity (p-value $<$ 0.01) are associated with a return of a voxel to intensity values closer to that of normal-appearing tissue. The second model is a function-on-scalar regression, which allows for assessment of the individual time points at which the covariates are associated with the profiles. In the function-on-scalar regression both age and distance to the boundary were found to have a statistically significant association with the profiles at some time point.  The methodology presented in this article shows promise for both understanding the mechanisms of tissue damage in the disease MS and may prove to be useful for evaluating the impact of treatments for the disease in clinical trials.

\section*{Keywords}

\noindent Structural magnetic resonance imaging; Multi-sequence imaging; Longitudinal study; Multiple sclerosis; Longitudinal lesion behavior; Principal component analysis regression; Function-on-scalar regression; Expert rater trial

\section*{Abbreviations}
\begin{itemize} 
\item Confidence Interval (CI)
\item Fluid attenuated inversion recovery (FLAIR)
\item National Institute of Neurological Disease and Stroke (NINDS)
\item Normal-appearing white matter (NAWM)
\item Magnetic resonance imaging (MRI)
\item Multiple sclerosis (MS)
\item Principal component (PC) 
\item Principal component analysis (PCA)
\item Proton density-weighted (PD)
\item Relapsing remitting MS (RRMS)
\item Secondary progressive MS (SPMS)
\item Standard deviation (sd)
\item T1-weighted (T1)
\item T2 -weighted (T2)
\item Tesla (T)
\end{itemize} 

\section{Introduction}

Structural magnetic resonance imaging (MRI) can be used to detect lesions in the brains of multiple sclerosis (MS) patients. The formation of these lesions is a complex process involving inflammation, tissue damage, and repair, all of which MRI has been shown to be sensitive to.  The McDonald criteria for diagnosis of MS have emphasized the key role of dissemination of lesions in the central nervous system on MRI not only in space, but also in time \citep{polman2011diagnostic}.  Characterizing the longitudinal behavior of lesions on structural MRI is therefore likely to be important for monitoring disease progression and response to therapy, and for understanding the etiology of the disease.   The clinico-radiological paradox refers to the poor association between clinical findings and radiological extent of involvement on MRI using traditional volumetric measures \citep{barkhof2002clinico}. Here we address this paradox by studying the association between the longitudinal behavior of lesions on structural MRI and clinical covariates.

Previous work to characterize the longitudinal behavior of lesions on structural MRI has only involved a single structural MRI sequence, proton density-weighted (PD) imaging \citep{ meier2003time, meier2006mri, meier2007time}. These approaches used a parametric model at the voxel level. Voxel-specific, biologically relevant features were then produced: hyperintensity above baseline, residual hyperintensity, and length of time at a given intensity level.   It was found that the maximal insult within a lesion occurred at the center of the lesion, that lower initial intensity within a lesion was predictive of repair, and that most lesion activity did not last beyond 10 weeks.  Some limitations of these studies were: (1) the MRI for these analysis consisted of frequent, bi-weekly scans; and (2) only one MRI sequence (PD) was used to characterize the behavior of the lesions, which ignores information known to be available in the other sequences \citep{mcfarland2002role}.  

Previous work designed to relate longitudinal changes in lesions on MRI and clinical covariates was recently published by \cite{ghassemi2014quantitative}.  The change over a 2-year period in normalized T1-weighted (T1) intensity within new T2-weighted (T2) lesions was compared in pediatric and adult-onset MS patients to investigate differences in the behaviors of lesions between the two populations. A generalized linear mixed-effects model was used to relate clinical covariates, such as disease duration and treatments, to changes in intensity in the MRI.  No statistically significant relationships were found, but a difference in the change in intensity between adults and pediatric MS patients was found, with pediatric patients showing more recovery than adults.   Work has also been done to relate longitudinal changes in lesion intensity to sample size calculations for clinical trials.  \cite{reich2015sample} used the change in the 25th percentile of intensity-normalized PD signal within a lesion over time to calculate sample sizes for clinical trials data for different lengths of trials.   The 25th, 50th, and 75th percentiles of multiple MRI sequences were assessed, and it was found that the 25th percentile of the normalized PD yielded the smallest sample size requirements.

In this study, we use MRI acquired from the National Institute of Neurological Disease and Stroke (NINDS), with subjects being scanned on average once every 37 days (sd 52.3, range [13, 889]) and an average  of 21 scans per subject (sd 8.0, range [10, 37]).  We first describe a pipeline for extracting the longitudinal profiles in lesion intensities from the T1, T2, T2-weighted fluid attenuated inversion recovery (FLAIR), and PD sequences.  Next, we describe two models for relating the longitudinal information from these profiles to clinical covariates.  The first model, a PCA regression model, collapses the information from the profiles into a single scalar.  This scalar value is found to identify areas of slow and persistent, long-term intensity changes.  This finding is validated by an expert rater trial with two raters, a neuroradiologist and a neurologist.  The scalar value is then related to clinical information in a mixed-model regression.  The second model, a function-on-scalar regression, allows for assessment of the time points at which the covariates are impacting the profiles. 


 \begin{figure}[h!]
   \begin{center}
 \includegraphics[width=4in]{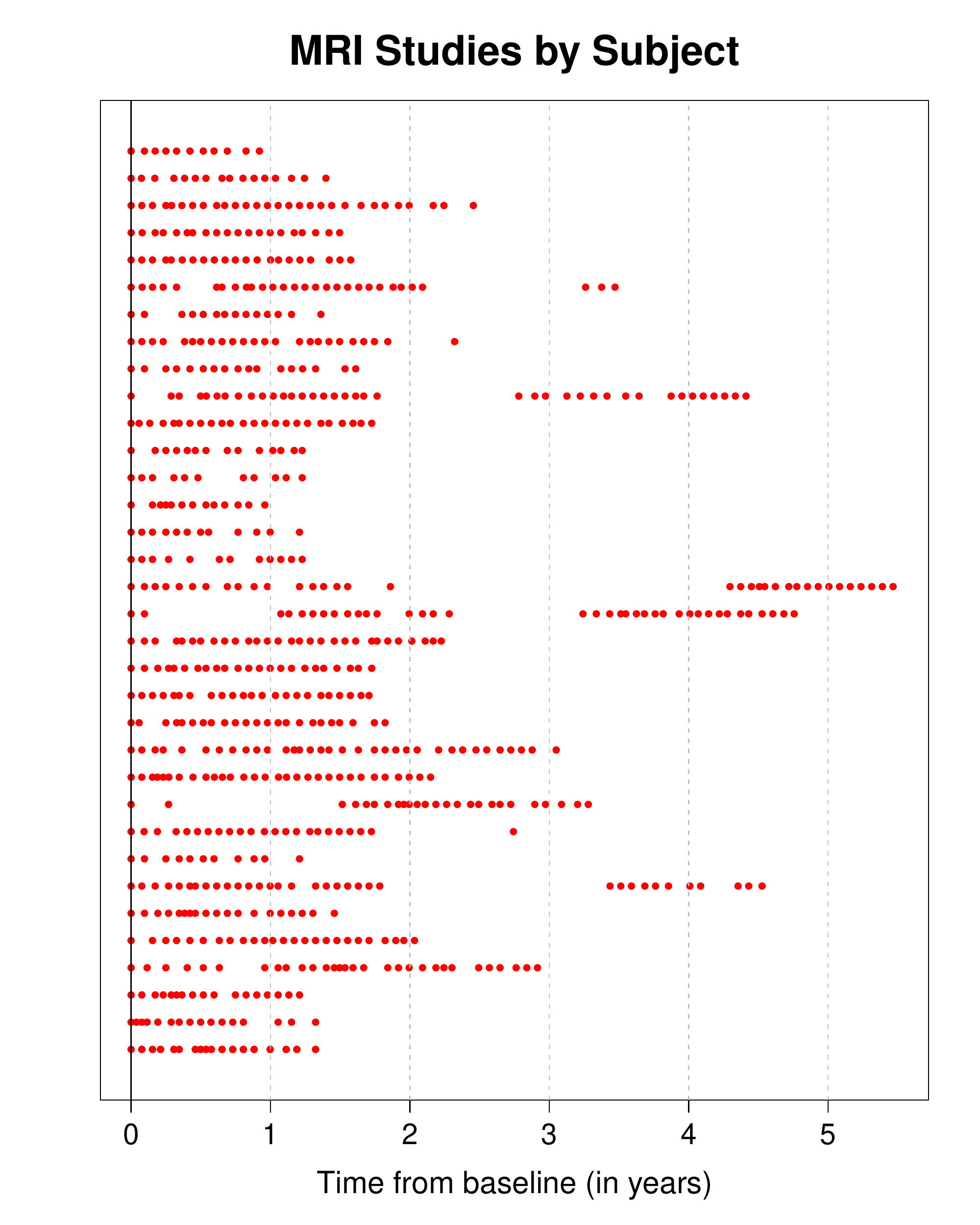}
 \end{center}
 \captionof{figure}{ \textbf{The time points at which each of the 34 subjects included in the analysis were scanned.}  Each row of the plot is a subject, and each point in the plot represents an MRI study.   The horizontal axis represents the time from the subject's baseline visit in years.  The subjects appear on the plot in random order.  Subjects had many scans, typically located at close intervals.  However, subjects were measured at different frequencies and some subjects had larger periods of time between observations. }
       \label{fig:scan_time}
   \end{figure}

\section{Material and Methods} 

In this section, we first describe the patient demographics, followed by the experimental methods, including the image acquisition and the image preprocessing.  Next, in the subsection \textit{Longitudinal Profile Pipeline}, we describe a pipeline for extracting the multi-sequence longitudinal profiles of the voxel intensities from the four sequences within lesion tissue.  In the subsection \textit{Modeling the Association with Clinical Covariates}, we describe two models that relate the clinical information to the four intensity profiles.  Last, in the subsection \textit{Expert Validation}, we describe a study to validate the lesion segmentation and the outcome from the first model.   All analysis, except for image preprocessing, was performed in the R environment \citep{R} using the R package oro.nifti \citep{oro}.

\subsection{Patient Demographics} 

For this analysis, we use 60 subjects scanned at the NINDS, with the earliest scan performed in 2000 and the most recent scan performed in 2008.  Three subjects were excluded  during the expert validation because it was found that the longitudinal registration had failed, causing overall poor segmentation of areas of abnormal signal intensity.  After exclusion of these subjects and subjects that did not have voxels that met a prespecified inclusion criteria, described in detail later in this section, there were 34 subjects left in the analysis.   The 34 subjects included in the analysis had an average of 21 scans each (sd 8.0, range [10, 37]).   Figure~\ref{fig:scan_time} shows the time points at which each of the 34 subjects were scanned.    Each row of the plot corresponds to a subject, and each point in the plot represents an MRI study, with time from the subject's baseline visit in years along the horizontal axis.  The total follow-up time per subject was on average 2.2 years (sd 1.2, range [0.9, 5.5]).  The mean age of the subjects at baseline was 37 years (sd 10.1, range [18,60]).    At baseline, there were 30 subjects with relapsing remitting MS (RRMS) and 4 subjects with secondary progressive MS (SPMS) .  There were 20 females and 14 males,14 subjects on disease-modifying treatment, and 2 subjects received steroids at the baseline visit.  The disease-modifying treatments and use of steroids for many of these subjects changed at subsequent follow-up visits. 

\subsection{Experimental Methods} 

\subsubsection{Image Acquisition and Preprocessing} 

Whole-brain 2D FLAIR, PD, T2, and 3D T1 volumes were acquired in a 1.5 tesla (T) MRI scanner (Signa Excite HDxt; GE Healthcare, Milwaukee, Wisconsin) using the body coil for transmission.  The 2D FLAIR, PD, and T2 volumes were acquired using fast-spin-echo sequences and the 3D T1 volume using a gradient-echo sequence. The PD and T2 volumes were acquired as short and long echoes from the same sequence. The scanning parameters were clinically optimized for each acquired image. 

For image preprocessing, we use Medical Image Processing Analysis and Visualization  (http:// mipav.cit.nih.gov) and the Java Image Science Toolkit (http://www.nitrc.org/projects/jist) \citep{lucas2010java}. We interpolate all images for each subject at each visit to a voxel size of $1$ $mm^3$ and rigidly align all of the T1 volumes to the Montreal Neurological Institute standard space \citep{fonov2009unbiased}. We rigidly align the T2, FLAIR, and PD image to the T1 for each subject at each visit.  We then find the average of the T1 volume for all the studies, for each subject, and register the longitudinal collection of the T1, T2, FLAIR, and PD images for the subject to this average image. We remove extracerebral voxels using a skull-stripping procedure \citep{carass2007joint}.  We automatically segment the entire brain using the T1 and FLAIR images \citep{shiee2010topology}.  After preprocessing, studies were manually quality controlled by a researcher with over three years experience with structural MRI (EMS).  Studies with motion or other artifacts were removed.  

\subsection{Longitudinal Profile Pipeline} 

The procedure for extracting the longitudinal voxel level lesion profiles is divided into four steps: (1)  identifying voxels with new lesion formation, (2) intensity normalization, (3) temporal alignment, and (4) temporal interpolation.  All voxels in this analysis are part of incident or enlarging lesions detected during the subject's follow-up period.  All voxels that are part of lesions that existed at baseline are excluded from the analysis. 

\begin{figure}[h!]
   \begin{center}
 \includegraphics[width=4.5in]{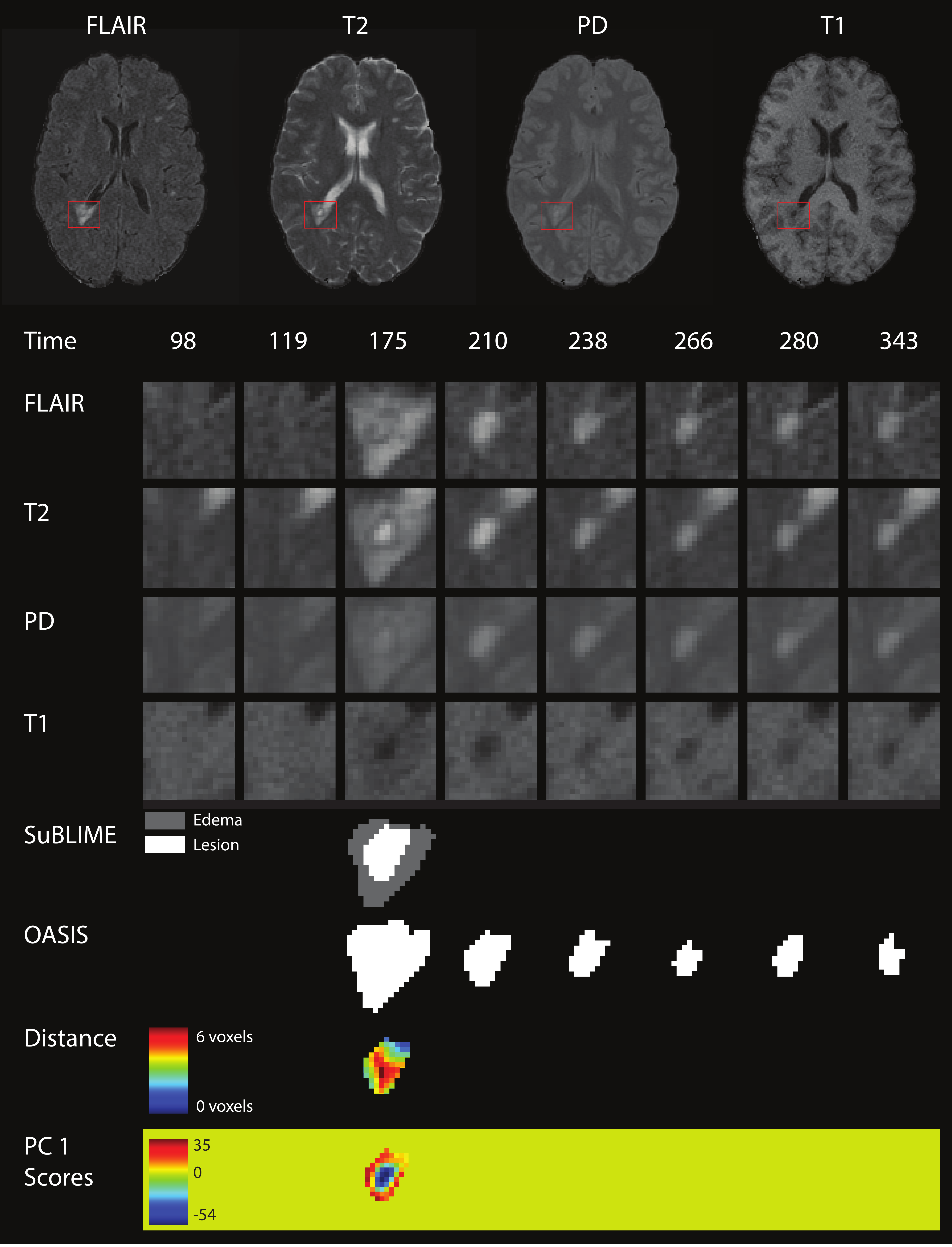}
 \end{center}
 \captionof{figure}{ \textbf{Longitudinal MRI studies within lesions.} The first row of the figure shows an axial slice from the multiple MRI sequences, 175 days after baseline (from left to right, the FLAIR, T2, PD, and T1 sequences).  In each sequence, a red box shows an area with a lesion that develops during the follow-up period.  In the subsequent rows of the figure, we show the longitudinal behavior within this red box.  Each column of the figure represents a different MRI study, starting at 98 days after baseline in the far left column and going until 343 days after baseline. A lesion is first identified in this area at 175 days. The first four rows show the longitudinal behavior of the FLAIR, T2, PD, and T1 sequences.  The next rows show the SuBLIME segmentation of lesion incidence for each study, the OASIS segmentation of lesion presence in each study, and a map of the distance to the boundary of the SuBLIME segmentation.  The SuBLIME segmentation has been further divided into areas of edema and lesion.  The last row shows the score on the first PC, which identifies areas of lesion repair and permanent damage.}
       \label{fig:longit}
   \end{figure}

Figure~\ref{fig:longit} shows an example of a lesion that is detected during a subject's follow-up period.  The first row of Figure~\ref{fig:longit} shows the multiple MRI sequences at one time point (from left to right, the FLAIR, T2, PD, and T1 sequences).  In each sequence, a red box shows an area with a lesion that develops during the follow-up period.  The subsequent 4 rows of the figure show the longitudinal behavior within this red box.  Each column of the figure shows a different MRI study, starting at 98 days after baseline in the far left column and going until 343 days after baseline.  The lesion in the red box is first seen 175 days after baseline.

\subsubsection{ Identifying Voxels with New Lesion Formation }  
When identifying voxels with new lesion formation, we distinguish between areas that contain vasogenic edema (which we will refer to simply as ``edema") and actual lesion, which both manifest as areas of abnormal signal intensity, especially on the T2-weighted sequences.  For this analysis, we are interested in areas with tissue damage, as opposed to edema.   To identify areas with new lesion formation, we first find areas in the MRI with new abnormal signal intensity, which includes both edema and lesion. We then segment lesions by analyzing subsequent visit data (details provided below).

SuBLIME segmentation of voxel level lesion incidence and enlargement is a method for detecting voxels that are part of an area of new abnormal signal intensity between two MRI studies \citep{sweeney2013automatic}.  For each subject, we produce SuBLIME maps between the respective sets of consecutive MRI studies.  We exclude all abnormal signal intensity areas that contained fewer than 27 voxels, as these areas could be artifact or noise.  We then produce cross-sectional lesion segmentations using OASIS segmentation of abnormal signal presence \citep{sweeney2013oasis}.  As the signal from edema disappears rapidly from the MRI after the lesion formation, we locate the incident abnormal signal voxels using SuBLIME, but only include the voxels that are detected by OASIS at the following study visit, as these voxels should not contain edema.  Therefore, only voxels that have an MRI study within 40 days after SuBLIME detects the area of abnormal signal intensity, where the intensity remains in the OASIS maps, are considered as lesion tissue and used in this analysis, as by this time edema would subside.  We use expert validation by a neuroradiologist and a neurologist, both with experience in MS imaging, to confirm that this method is identifying lesion tissue, which we describe in detail in the subsection \textit{Expert Validation}.  Figure~\ref{fig:longit} shows the SuBLIME segmentation for each study, the OASIS segmentation for each study, and a map of the distance to the boundary of the SuBLIME segmentation.  The row corresponding to the SuBLIME segmentation is further divided into edema and lesion voxels using the method described above.  Only voxels that are part of lesion tissue are used in the analysis.

\subsubsection{ Intensity Normalization }  
Structural MRI is acquired in arbitrary units.  Therefore, in addition to pulse sequence similarity, intensity normalization is paramount for comparing intensities in a voxel over time within subject and for comparing voxel intensities between subjects.   We normalize each sequence separately on each scan by calculating the mean and standard deviation over a mask of the normal-appearing white matter (NAWM) from the brain segmentation described in the subsection \textit{Image Acquisition and Preprocessing} \citep{shiee2010topology}.  We then subtract the mean from the intensity in each voxel and divide by the standard deviation \citep{shinohara2011population, shinohara2014statistical}.    Let $S_{ilv} (t)$ be the observed intensity from imaging sequence $S$ in voxel $v$ for subject $i$ in lesion $l$ at study time $t$, with $S = $ FLAIR, T1, T2, and PD.  Let $\mu_{Si} (t) $ and $\sigma_{Si} (t) $ be the mean and standard deviation, respectively, over the NAWM mask for sequence $S$ at scan time $t$ for subject $i$ .  Then the normalized intensity in voxel $v$ in lesion $l$ for subject $i$ at scan time $t$ is:

\begin{equation}
S^N_{ilv} (t)   = \frac{S_{ilv}(t) - \mu_{Si}(t)  }{ \sigma_{Si} (t) } \nonumber 
\end{equation}

\noindent Thus, all image intensities are expressed as a departure, in multiples of standard deviation of white matter intensities, from the subject's mean normal-appearing white matter (NAWM) in each imaging sequence. 

\begin{figure}[h!]
   \begin{center}
 \includegraphics[width=6in]{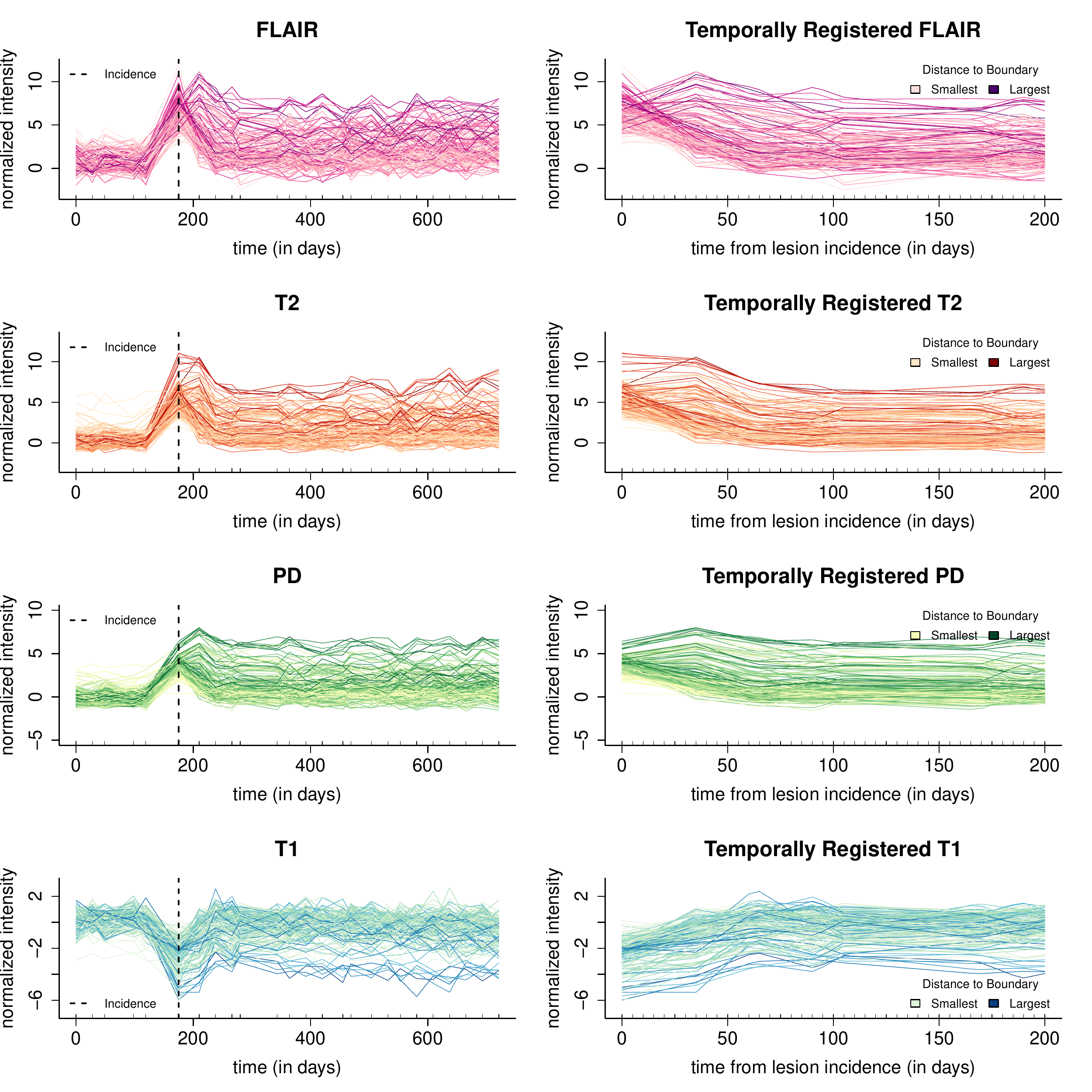}
 \end{center}
 \captionof{figure}{ \textbf{Multi-sequence lesion profiles.} The first column of the figure shows the full longitudinal profiles from all four sequences (from top to bottom, the FLAIR, T2, PD, and T1 sequences).  The profiles are from 150 randomly sampled voxels from the lesion in Figure~\ref{fig:longit}, and for display purposes the periods between each study have been linearly interpolated.   Each line in the plot represents the longitudinal profile from a single voxel.  The x-axis shows the time in days from the subjects's baseline visit,  the point of lesion incidence is denoted by a dashed line, and the y-axis shows the normalized sequence intensities.   The second column shows the same voxels after temporal alignment and linear interpolation over the 200 day period after incidence. The profiles are colored by distance to the boundary of abnormal MRI signal. }
       \label{fig:trajectories}
   \end{figure}

The normalized intensity longitudinal profiles from the lesion in Figure~\ref{fig:longit} for all four sequences can be seen in the first column of Figure~\ref{fig:trajectories}.  From top to bottom in  the first column of Figure~\ref{fig:trajectories} we have the profiles from 150 randomly sampled voxels from the lesion in Figure~\ref{fig:longit}  for the FLAIR, T2, PD, and T1 sequences.   Each line in the plot represents the longitudinal profile from a single voxel.  The x-axis shows the time in days from the baseline visit,  with the point of lesion incidence denoted by a vertical dashed line, and the y-axis shows the normalized sequence intensities.

\subsubsection{ Temporal Alignment }  

The date of the study visit at which SuBLIME detects the lesion voxels is considered the time of incidence for this voxel.  If a voxel is determined to be a new or enlarging lesion by SuBLIME more than once over the follow-up time, the first occurrence is considered to be the time of lesion incidence for that voxel.  Voxel profiles from incident lesions during the follow-up of each subject are aligned in time, using the time of incidence as time 0, therefore any observations before incidence have a negative time and after lesion incidence have a positive time.   Let $t'$ denote this aligned time scale.  Then we have $S^N_{ilv}(t')$, where $S^N_{ilv}(0)$ indicates the intensity in sequence $S$ at the time of lesion incidence.    

\subsubsection{ Temporal Interpolation}  

Next we perform a temporal linear interpolation so that all voxels are observed on the same time grid.  In this work, we are interested in the lesion dynamics only after lesion incidence, therefore we perform the linear interpolation within the window after lesion incidence and up to 200 days post-incidence.  The end point of 200 days is selected as it has been previously found that new T2 lesions show the most dramatic changes in intensity for three to four months \citep{meier2007time}, and we opt to be conservative and include data beyond this reported stabilization point.   Voxels are selected for the analysis if the subject has at least one visit 200 days or more after lesion incidence.   Of the 60 subjects in this analysis, 34 have voxel profiles meeting this inclusion criteria, after removing the three subjects for poor longitudinal registration.   We linearly interpolate over a grid of 0 to 200 days by increments of 5 days so that all profiles are observed on the same time grid.   We denote the vector of observations from a voxel over this time grid for sequence $S$ as $S^N_{ilv}$, where $S^N_{ilv}$ is a $1 \times 41$ vector.   The second column of Figure~\ref{fig:trajectories} shows the temporally registered and linearly interpolated profiles over the period of 0 to 200 days for the lesion in Figure~\ref{fig:longit} for the same 150 randomly sampled voxels as shown in the first column.

\subsection{Modeling the Association with Clinical Covariates} 

We propose two models for relating the information in the lesion profiles to the clinical covariates.  The clinical information for each subject at each study visit is MS disease subtype,  age, sex, an indicator of treatment with steroids, an indicator of disease-modifying treatment, and distance to the boundary of an  area of abnormal signal intensity.   An example of distance to the boundary of an area of abnormal signal intensity for the lesion in  Figure~\ref{fig:longit} can be seen in the eighth row of the Figure~\ref{fig:longit}.    We center age at the mean age over all of the voxel level observations of 36 years.  During the observation period, many of the subjects were enrolled in clinical trials at NINDS to test various experimental therapies.   Our indicator of disease-modifying treatment indicates treatment with any of the Food and Drug Administration approved treatments interferon beta 1-a (intramuscular or subcutaneous), interferon beta 1-b, and glatiramer acetate, as well as experimental therapy.  As many of the covariates change over time, we model the relationship between the lesion profiles  and the value of the covariate at the time of lesion incidence for the particular profiles.   For the following analysis, we have a total of  57,908 voxels from 315 lesions in 34 subjects. 

\subsubsection{Principal Component Analysis and Regression}

To perform PCA, we first concatenate the voxel profiles for each voxel from the four sequences together.  For each sequence and at each voxel there is a $1 \times 41$ vector of longitudinal intensities, $S^N_{i l v}$.  Let $I_{ilv}$ denote the $1 \times 164$ dimensional vector of the four concatenated profiles, $S^N_{i l v}$, from subject $i$ lesion $l$ and voxel $v$.  More precisely, 

\begin{equation}
I_{ilv}  = \{ FLAIR^N_{ilv},  T1^N_{ilv}, T2^N_{ilv},  PD^N_{ilv} \} 
\end{equation}

\noindent and we index the entries of $I_{ilv}(j)$, where $j=1,\ldots,164$ is the $j^{th}$ entry of the concatenated vector. Note that we first remove the mean from the concatenated profiles and then perform a PCA on these concatenated profiles.   Let  $\phi_k$ denote the $k^{th}$ PC, where  $\phi_k$ is also indexed by $j$.  The relationship between the score on the $k^{th}$ PC, $\xi_{ilv}(k)$, a scalar,  and the observed trajectory for $I_{ilv} \left( j \right)$ is:

\begin{equation} 
I_{ilv} \left( j \right) =\sum_{k=1}^K \xi_{ilv}(k) \phi_k  \left( j \right) , 
\end{equation} 

\noindent with $\xi_{ilv}(k)$ being the scalar score on the $k^{th}$ eigenfunction. We will focus on the first PC, $\phi_1$, and the score on this component, $\xi_{ilv}(1) $.   The first PC is found to identify slow, long-termintensity changes at the voxel level within lesions, with positive PC scores corresponding to a return of the voxel to intensity values closer to that of normal-appearing tissue and negative scores corresponding to the voxels maintaining intensity values closer to those at lesion incidence.  We use expert validation from a neuroradiologist and a neurologist to confirm this finding, with details provided in the subsection \textit{Expert Validation}.    The score on the first PC, $\xi_{ilv}(1)$, collapses the full profiles at each voxel from the four sequences into a single scalar.  We use only the score on the first PC in this analysis, as the other PCs explain only $25\%$ of the variation in the data and were not found to identify any biological processes. To assess variability in both the mean and the first PC, we bootstrap this procedure by resampling subjects with replacement 1000 times \citep{efron1994introduction}.

We now introduce a linear mixed-effects model to relate the score on the first PC to the clinical covariates \citep{mcculloch2001generalized}. We use the value of the covariate at the time of lesion incidence for the particular profiles, which can vary within subject. Thus, for added precision, the covariates that change over time are indexed by the subject index $i$, lesion index $l$ and voxel index $v$, as voxels from the same lesion may have different times of incidence. For example, the sex of the subject does not change by time of lesion incidence, so it is only indexed by $i$.  In contrast, age of the subject changes with lesion incidence and is indexed by $i$, $l$ and $v$.  We also add random effects for subject and lesion, which we denote by $b_i$ and $b_l$, respectively, with both following a normal distribution:  $b_i \sim N \left(0, \sigma_i^2 \right)$ and $b_l \sim N \left(0, \sigma_l^2 \right)$, where $\sigma^2$ denotes the variance of the random effects. We consider the following basic model for the association between $\xi_{ilv}(1)$ and the covariates:

\begin{align} 
\xi_{ilv}(1)  & = \beta_0 + \beta_1 \text{SPMS}_{ilv}  +  \beta_{2} \text{ Distance} _{ilv} + \beta_3 \text{Age}_{ilv} + \beta_4 \text{ (Age $-$ 4)}_{+ ilv}   \nonumber \\
&  + \beta_{5} \text{Steroids}_{ilv} +   \beta_{6} \text{Male}_{i}  + \beta_{7} \text{Treatment}_{ilv}  + b_i + b_l + \epsilon_{ilv}  \nonumber 
\end{align} 

\noindent We assume that the error terms are independent identically distributed, with each $\epsilon_{ilv}$ following a normal distribution, $\epsilon_{ilv} \sim N \left(0,  \sigma_{\epsilon}^2\right) $.  In the model, the term SPMS is an indicator of having SPMS where the comparison group is RRMS.  Note that the age term has been centered at the mean age of 36 years.  The term $\text{ (Age $-$ 4)}_{+ ilv} = \text{Age}_{ilv} \cdot 1( \text{Age}_{ilv}>4)$ is a spline term for centered age over 4 years (or age over 40 years), which was included in the model after visualizing the relationship between the scores on the first PC and age.  This visualization can be found in the Appendix.  We also investigated simpler models with the same mixed-effects structure, but where we considered each covariate separately.

To test for associations, we use two procedures. First, we perform a parametric bootstrapping procedure \citep{efron1994introduction}, and second we calculate p-values using a normal approximation for the distribution of the fixed-effects in the mixed-effects model \citep{barr2013random}.  We use 1000 bootstrap samples  for the bootstrap procedure.  We perform the parametric bootstrap because steroid use and disease subtype of SPMS did not always appear in the nonparametric bootstrap sample.  A complete description of the parametric bootstrap procedure is found in the Appendix.   We also use the normal approximation, as this approximation has been found to be a reasonable approximation for the distribution of the fixed-effects in most settings \citep{barr2013random}. 

\subsection{Expert Validation}
 
 We use expert validation to determine the quality of the lesion segmentation (excluding edema tissue) and the score on the first PC for identifying areas of slow, long-term intensity change.   For the validation we use two raters, a neuroradiologist with 11 years of experience with research in MS (DSR) and a neurologist with 4 years of experience with research in MS (MKS).  For each lesion, we first determine the axial slice of the image that contains the largest number of voxels with abnormal signal intensity.  Then for each lesion the two raters are presented the following: (1) the full axial slice for the FLAIR,  T2, PD, and T1 volumes that contains the largest number of voxels with abnormal signal intensity; (2) the entire collection of longitudinal scans for a box containing the abnormal signal intensity in the FLAIR, T2, PD, and T1 volumes for this axial slice; (3) the segmentation of the lesion and edema tissue within this box; (4) the score on the first PC for the voxels segmented as lesion tissue within this box and a scale for the intensities within this image; (5) the entire collection of longitudinal scans for the FLAIR, T2, PD, and T1 weighted volumes within this box with the score for the first PC overlaid on the images for each scan after lesion incidence.    The raters are then asked to rate the quality of the lesion tissue segmentation and the score on the first PC for identifying areas of slow, long-termintensity changes on an integer scale from 1 to 4, with each rating corresponding to the following: (1) failed miserably; (2) some redeeming features; (3) passed with minor errors; and (4) passed.  Examples of the images presented to the raters for each lesion that received a rating of 1,2,3, and 4 for the score on the first PC by both raters are provided in the Appendix. Forty-seven lesions are selected at random to be repeated in the analysis to assess intra-rater reliability.  
 
We report the median of the ratings of the lesion segmentation and the score on the first PC for each rater over all lesions.  To assess between-rater reliability, we report the Coehn's kappa coefficients  for the two raters separately, for both the ratings, over all of the lesions.  We also assess the within-rater agreement by reporting the kappa coefficient for the set of repeated lesions for each rater, for both the ratings.  We also report the kappa coefficient for the rating of the lesion segmentation and the score on the first PC for all lesions, for each rater, to determine if the quality of the segmentation and the quality of the score on the first PC are related.   We nonparametrically bootstrap with replacement the subjects 1000 times to produce the confidence intervals for the median of the ratings for each rater and the kappa coefficients for within- and between-rater agreement as well as the relationship between the quality of the segmentation and the score on the first PC.

\subsubsection{Function-on-Scalar Regression} 

The previous model is an attempt to collapse the information from the four profiles (across sequences) for a voxel into a single scalar at each voxel.  As an alternative, we also fit a function-on-scalar regression model \citep{reiss2010fast}, where we can investigate the relationship between the covariates of interest and the profile at each time point.  We fit a function-on-scalar regression model for each sequence separately.  For simplicity of notation, we now use $t$ for the registered time, as opposed to $t'$.  The outcome in the model is the full lesion intensity profile:
\begin{align} 
 S^N_{i l v}(t)  & = \beta'_0(t) + \beta_1'(t)  \text{SPMS}_{ilv} +  \beta_{2}'(t) \text{ Distance} _{ilv} + \beta_3'(t) \text{Age}_{ilv}   + \beta_4'(t) \text{ (Age $-$ 4 )}_{+ ilv}    \nonumber \\
&   + \beta_{5}'(t) \text{Steroids}_{ilv} +   \beta_{6}'(t) \text{Male}_{i}  + \beta_{7}'(t) \text{Treatment}_{ilv} + \epsilon_{ilv} \left( t \right) \nonumber 
\end{align} 

\noindent for  $S = $ FLAIR, T1, T2, and PD.  To fit the model, we use a two-step function-on-scalar regression implemented in the R package refund \citep{refund}.   The procedure first fits a scalar-on-scalar regression at each individual time point.  Then the resulting coefficient functions are smoothed over time using a cubic spline basis with a penalty on the second derivative. 

To assess the variability in the coefficient functions and provide bootstrapped ,point wise $95\%$  confidence intervals, we non-parametrically bootstrap subjects 1000 times.   When samples do not contain subjects with a covariate, for example the indicator of steroids, we remove this sample from the bootstrap and replace it with another sample.    The difference between the function-on-scalar regression and the PCA regression model is that PCA collapses the entire temporal intensity profile of the voxel into a scalar. By contrast the function-on-scalar regression investigates the association at every time point. While function-on-scalar regression is more comprehensive and interpretable, it is more appropriate when there are strong functional effects that are not captured by one or two principal components.


\section{Results} 

\subsection{Principal Component Analysis and Regression} 

\begin{figure}[h!]
   \begin{center}
 \includegraphics[width=6in]{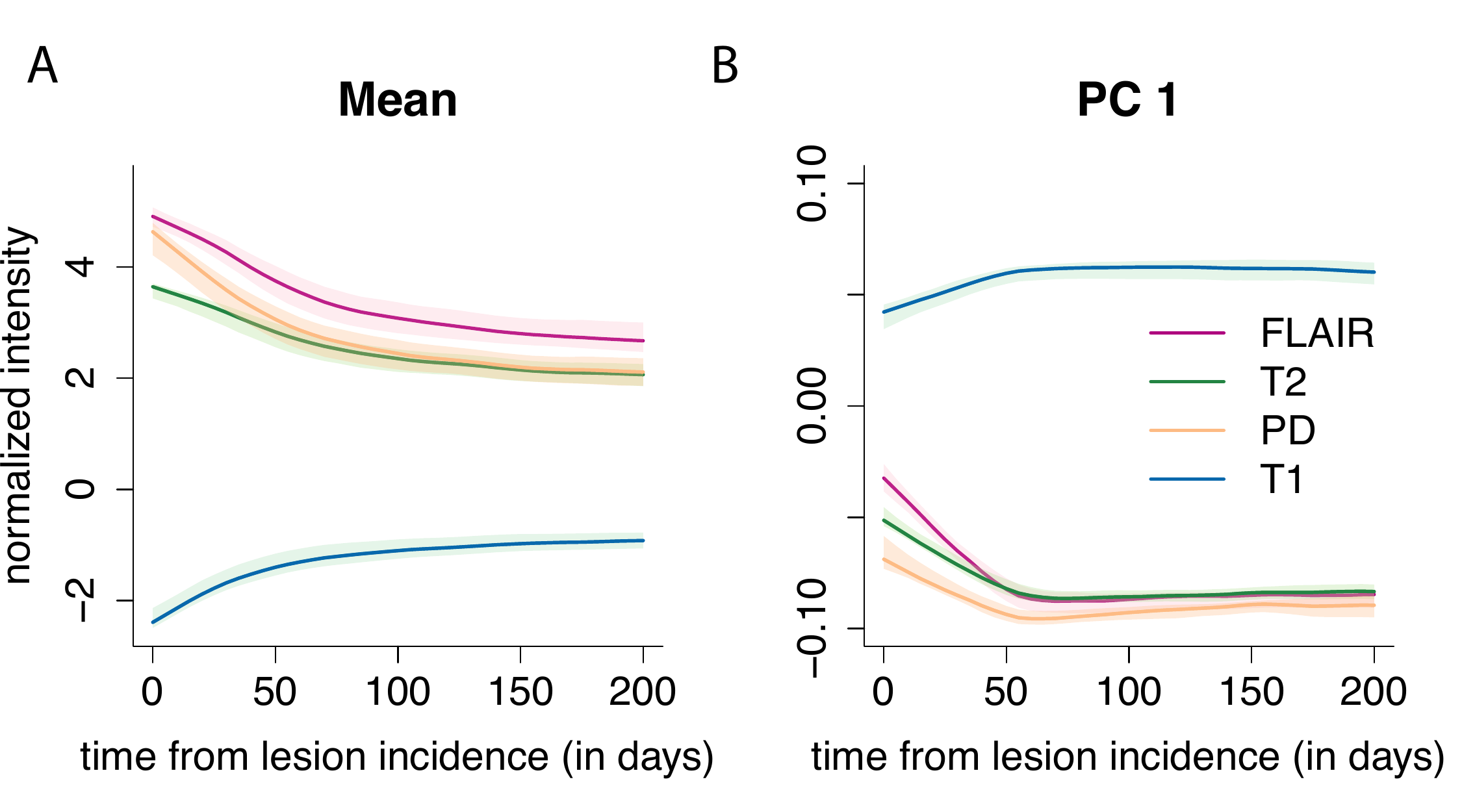}
 \end{center}
 \captionof{figure}{ \textbf{The mean profile and first PC for each of the four sequences.}  Panel A of the figure shows the mean profiles for each of the imaging sequences over the registered 200 day period, and panel B shows the first PC for each of the imaging sequences.  The first PC explains 79\% of the variation in the concatenated longitudinal profiles.  Along the x-axis for both plots is plotted the time in days since lesion detection.  The $95\%$ confidence intervals in both panels are obtained using 1000 nonparametric bootstrap samples of subjects.}
       \label{fig:PCload}
   \end{figure}

In Figure~\ref{fig:PCload} A we show the mean profiles for each sequence over the registered 200 day period, and in Figure~\ref{fig:PCload} B we show the first PC for each sequence over the registered 200 day period.  The first PC explains 75\% (95\% CI: [72\%, 76\%]) of the variation in the concatenated longitudinal profiles.  The first PC,  $\phi_1$, is split into the different sequences for purposes of presentation.  The subfigures for both the mean and  the first PC show the bootstrapped $95\%$ confidence intervals.  Recall that the normalization procedure puts the volumes into units of standard deviations above the mean of the NAWM.  Therefore a value of 0 on the image corresponds to the average value of NAWM from the particular MRI scan. The mean profiles for the FLAIR, T2 and PD are all above 0 throughout the time course, as lesions are hyperintense on these sequences.   In contrast, the mean profile for the T1 sequence is below 0, as lesions are hypointense on this sequence.  The first PC for the FLAIR, T2 and PD is negative throughout the time course, with values closer to 0 at lesion incidence (time 0).  Positive scores on this PC indicate a decrease in the signal in these sequences, which corresponds to a return of the voxel to intensity values closer to that of normal-appearing tissue.  In contrast, negative scores indicate the voxel maintaining intensity values closer to those at lesion incidence, with more hypointensity than the average profile.   Similarly,  for T1 the first PC is positive throughout the time course, with values closer to 0 at lesion incidence.  Positive scores on this PC indicate increased signal on the T1.  As lesions are hypointense on the T1, this also indicates a return of the voxel to intensity values closer to that of normal-appearing tissue.  Negative scores again correspond to the voxels maintaining intensity values closer to those at lesion incidence.   Therefore, we consider the score on the first PC to capture slow, long-termintensity changes within the lesion at a voxel level.   In the last row of Figure~\ref{fig:longit} we see the PC scores from the lesion that is shown in the figure.  Here we see that the positive scores indicate areas of the lesion that return to values of normal-appearing tissue, while the negative scores show areas that remain at the intensity values at lesion incidence. 

   \begin{figure}[h!]
   \begin{center}
 \includegraphics[width=6in]{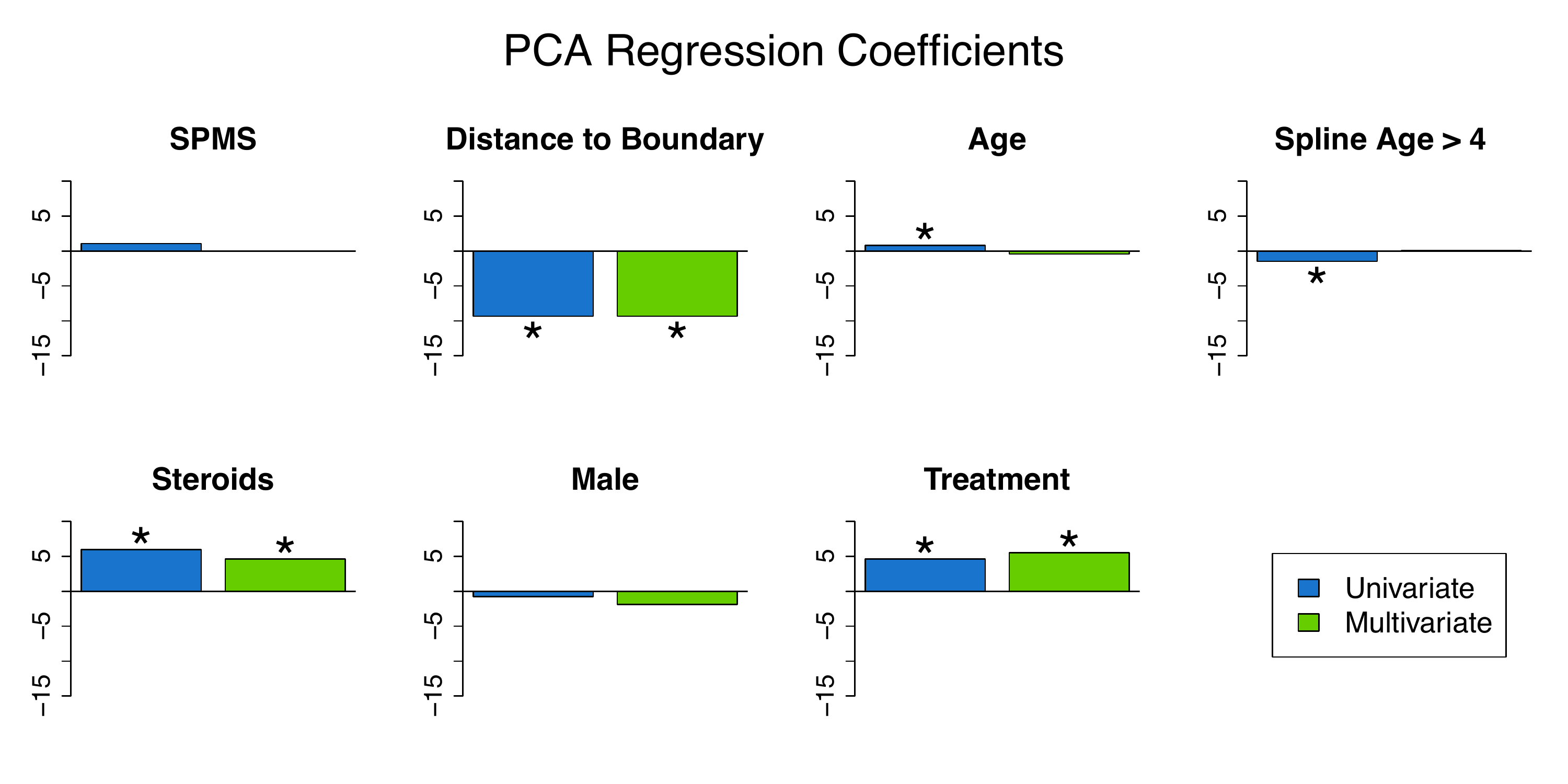}
 \end{center}
 \captionof{figure}{ \textbf{Coefficients from the PCA Regression model.} This figure shows bar plots of the coefficient estimates from the univariate and multivariate mixed-effects models with the score on the first PC as an outcome.  The results from the univariate model are shown in blue, and the results from the multivariate model are shown in green.   Asterisks indicate significance at the $5 \%$ level.  In both the univariate and multivariate models, disease-modifying therapy, steroids, and age were found to be significant. }
       \label{fig:coefficient}
   \end{figure}

In the second-stage analysis, we fit both univariate and multivariate mixed-effects models to investigate the relationship between the covariates and the score on the first PC.    The estimates of the coefficients from both models are shown in the bar plots in Figure~\ref{fig:coefficient}, with asterisks indicating statistical significance at the $5\%$ level using the bootstrapped 95\% confidence intervals.  Tables containing the coefficient estimates, standard errors, t-statistics, p-values using the normal approximation, and 95\% bootstrapped confidence intervals can be found in the Appendix for both the univariate and the multivariate models.  There are no differences in the conclusions determined by the normal approximation and the bootstrapped 95\% confidence intervals. For continuous covariates, such as age, the coefficient is interpreted as the expected change in the score on the first PC for a one unit increase in the covariate.  For binary variables, such as disease subtype, the coefficient is interpreted as the difference in the expected change in the first PC in the specified group.  Therefore,  positive coefficients are indicative of the voxel returning to intensity values closer to normal-appearing tissue with an increase in the covariate, while negative coefficients are indicative of the voxel maintaining the intensities at lesion incidence with an increase in the covariate (or in some rare cases having intensities that have an increasing departure from those of normal-appearing tissue over time with an increase in the covariate).   The results indicate that voxels that are farther away from the boundary have increased risk for maintaining abnormal signal intensity.  In this model, the coefficient for distance to the boundary has a value of  -9.39 (95\% CI:  [-9.56, -9.25] ), indicating that for a one voxel (or $1$ $mm$) increase in distance away from the boundary (toward the center of the lesion) the average value of the first PC decreases by  9.39, adjusting for the other coefficients and the random effects.   In the last row of Figure~\ref{fig:longit}, we see this spatial relationship between the score on the first PC and the distance to the lesion boundary; with positive scores near the boundary and negative scores near the center of the lesion.    In both models, we found the use of disease-modifying treatment and steroids to be associated with return of a voxel to the value of normal-appearing tissue.  The coefficient for treatment has a value of 5.39 (95\% CI:  [4.67, 6.08]), indicating that when subjects are on treatment the average value of the first PC increases by 5.39, adjusting for the other coefficients and the random effects.  The use of steroids has a similar interpretation, with a coefficient value of 4.26 (95\% CI:  [2.67, 5.85]).  In the Appendix, density plots for the score on the first PC by disease subtype, treatment with steroids, disease-modifying treatment, and sex are provided.

\subsection{Expert Validation} 
\begin{figure}[h!]
   \begin{center}
 \includegraphics[width=6in]{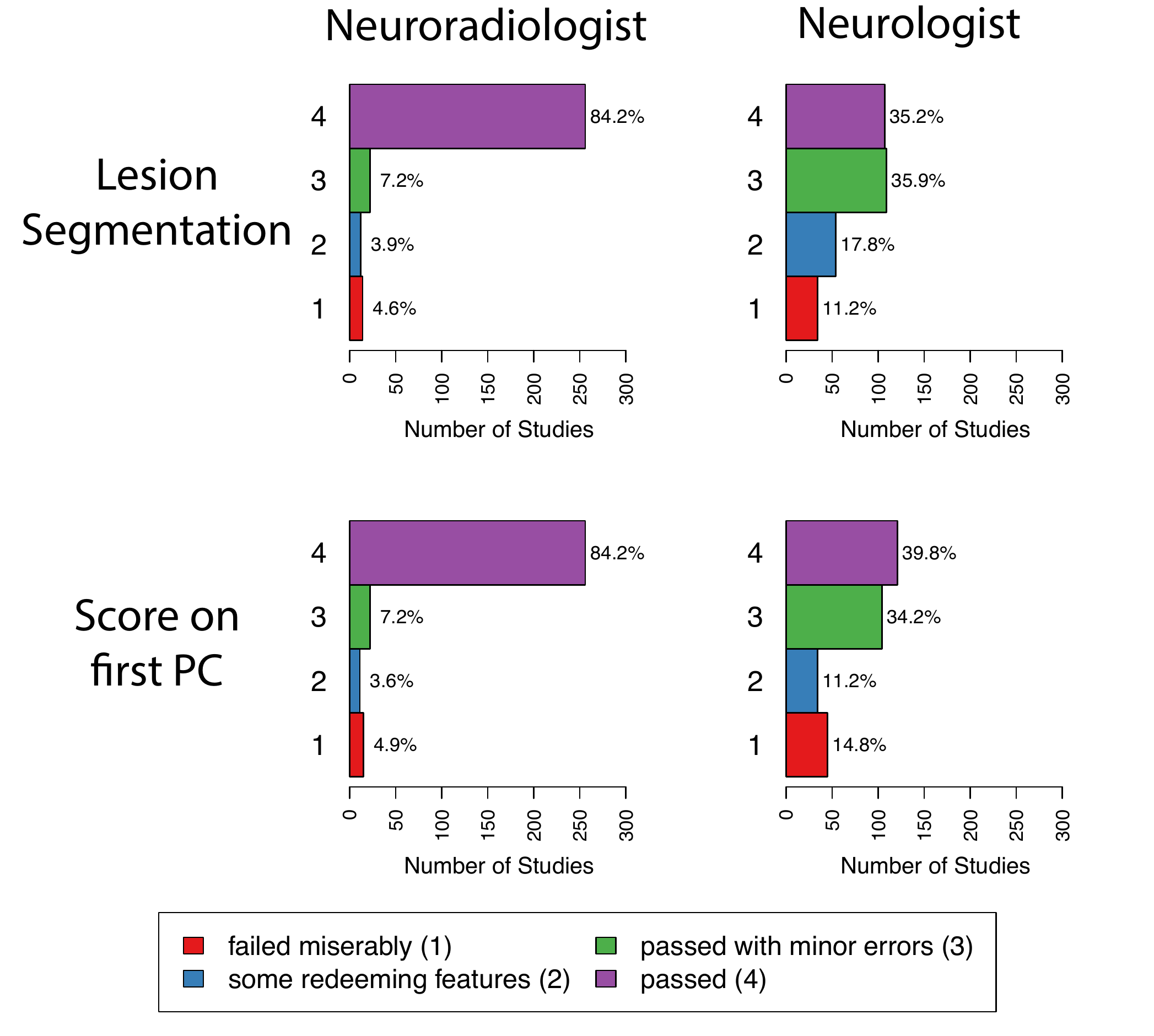}
 \end{center}
 \captionof{figure}{ \textbf{Distributions of the ratings for the two raters.}  The first row of plots shows the distributions of the ratings for the lesion segmentation, and the second row shows the ratings for the first PC.   Plots in the left column are ratings by the neuroradiologist, and plots on the right column are ratings by the neurologist.  Each plot shows the number of studies that failed miserably (1), had some redeeming features (2), passed with minor errors (3), and passed (4) along with the percentage of each rating. }
       \label{fig:ratings}
   \end{figure}

 Expert validation is used to determine the quality of the lesion segmentation (excluding edema tissue) and the score on the first PC for identifying areas of slow, long-termintensity change.   The distributions of the ratings for the two raters for both the lesion segmentation and the score on the first PC are shown in Figure~\ref{fig:ratings}.  The first row of plots in Figure~\ref{fig:ratings} shows the distribution of the ratings for the lesion segmentation and the second row shows the ratings for the score on the first PC.  Plots in the left column are ratings by the neuroradiologist, and plots on the right column are ratings by the neurologist.   The median rating for both the lesion segmentation and the first PC by the neuroradioloist are 4 (95\% CI: [4,4]), which is a rating of passed, the highest possible rating.  The median rating for both the lesion segmentation and the first PC by the neurologist are 3 (95\% CI: [3,3]), which is a rating of passed with minor errors.  Note that criteria for assigning scores were not discussed between the two raters prior to their respective analyses.

The Cohen's kappa coefficients for the within- and between-rater agreement for both the lesion segmentation and the score on the first PC are shown in Table~\ref{table:pearson}.  The values for the kappa coefficient range between $0$ and $1$, with a value of $1$ indicating total agreement and $0$ indicating no agreement. The within-rater agreement for the lesion segmentation and the score on the first PC are higher for the neuroradiologist than the neurologist.  There is only modest agreement between the neuroradiologist and neurologist on both ratings, with a kappa coefficient of 0.29 (95\% CI: [0.18, 0.41]) for the lesion segmentation and  0.24 (95\% CI: 0.11, 0.39) for the score on the first PC.  This is due, in part, to the fact that the neurologist spread ratings of the studies between 3 and 4, while the neuroradiologist gave more ratings of 4. 

The kappa coefficient for the agreement between the rating of the lesion segmentation and the score on the first PC is 0.97 (95\% CI: 0.93, 1.00)  for the neuroradiologist and 0.68 (95\% CI: 0.58,  0.78) for the neurologist.  The high correlation between these ratings, especially for the neuroradiologist, indicates that the quality of the segmentation impacts the quality of the score on the first PC.  Comments from the raters mirrored this finding, as many of the low scores for both the lesion segmentation and the first PC were due to: (1) missing the first time point of lesion incidence and segmenting it as new lesion at a later time point; (2) not segmenting the entire lesion; and (3) parts of the same lesion being segmented (unnecessarily) at different time points.  As both the ratings for the lesion segmentation and the score on the first PC were high, the quality of the lesion segmentation does not appear to be negatively impacting the method.

  \begin{table}
\centering

\begin{minipage}{.44\textwidth}
  \begin{tabular}{ |l|l|l| }
\hline
\multicolumn{3}{ |c|}{Lesion Segmentation} \\
\hline 
  & Neuroradiologist & Neurologist \\ \hline
 Neuroradiologist & 0.92; (0.76,0.99)  & 0.29; (0.18, 0.41) \\ \hline
Neurologist   &  &  0.75; (0.62, 0.86)   \\
\hline
\end{tabular}
\end{minipage}\hfill
\begin{minipage}{.443\textwidth}
  \begin{tabular}{ |l|l|l| }
\hline
\multicolumn{2}{ |c|}{Score on the First PC} \\
\hline 
 Neuroradiologist & Neurologist \\ \hline
 0.92; (0.76, 0.99)  &   0.24; (0.11, 0.39) \\ \hline
  & 0.72; (0.51, 0.86) \\
\hline
\end{tabular}
\end{minipage}\hfill
\caption{\textbf{Cohen's kappa coefficients for the ratings of the lesion segmentation and the score on the first PC.} The table on the left shows the kappa coefficients for the lesion segmentation, and the table on the right shows the same for the score on the first PC.  The between-rater agreement is reported using all lesions.  The within-rater agreement is reported using only the forty-seven repeated lesions. }
       \label{table:pearson}
\end{table}

\subsection{Function-on-Scalar Regression}

The resulting coefficient functions from the function-on-scalar regression with bootstrapped, point wise $95\%$ confidence intervals with the FLAIR profile as the outcome are shown in Figure~\ref{fig:fosrflair}.  Similar figures for the T2, PD, and T1 profiles as outcome are provided in the Appendix.  The coefficient functions for continuous variables in the function-on-scalar regression model are interpreted as the change in the expected profile at each time point for a one unit increase in the covariate.  Similarly, for binary variables, the coefficient function is interpreted as the change in the expected profile for the specified group.  For the FLAIR profiles, the coefficient functions corresponding to distance to the boundary and age have bootstrapped  $95\%$ confidence intervals that do not overlap with 0 across any of the time points, and are therefore statistically significant at the $.05$ level.  The coefficient function for distance to the boundary is greater than 0 throughout the entire trajectory, indicating that the further away from the boundary the voxel is, the more the FLAIR hyperintensity is maintained within the voxel.  For a one voxel (or $1$ $mm$) increase in distance away from the boundary (toward the center of the lesion) the average normalized intensity of the trajectory increases by around $0.5$ at all time points, adjusting for the other coefficients and the random effects.   The result for distance from the boundary agrees with the results from the PCA regression model. 

\begin{figure}[h!]
   \begin{center}
 \includegraphics[width=6in]{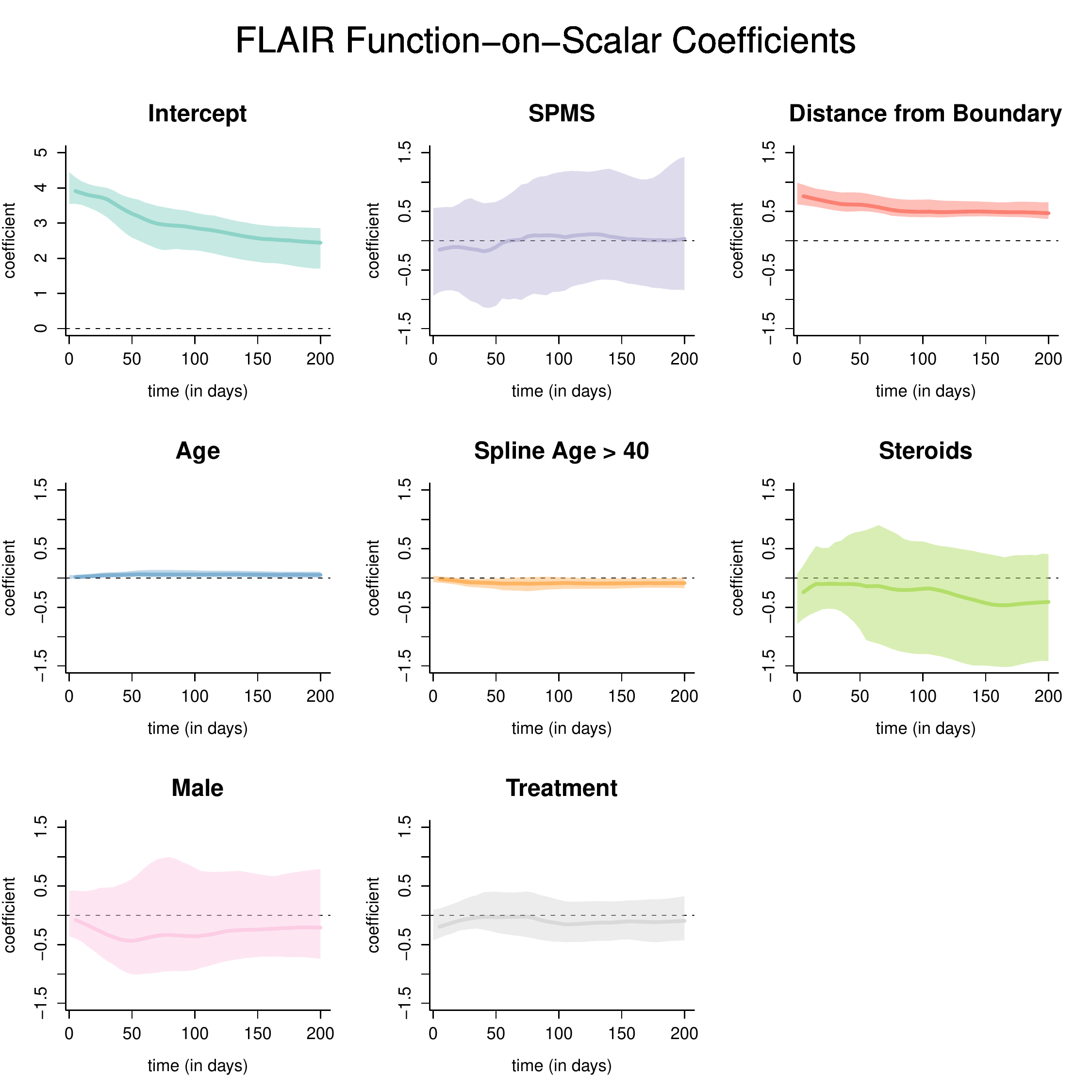}
 \end{center}
 \captionof{figure}{ \textbf{Coefficient functions from the function-on-scalar regression with the FLAIR profile as an outcome.}  Each dark line represents the coefficient function, and the shaded area represents a bootstrapped, point wise 95\% confidence interval.   Along the x-axis of each plot is the time in days from lesion incidence.  Along the y-axis is the value of the coefficient function at each time point. Only distance from the boundary and age were found to be different for 0 at any point along the profile. }
       \label{fig:fosrflair}
   \end{figure}

\section*{Discussion} 
 
We describe a pipeline to extract voxel level, multi-sequence longitudinal profiles from the FLAIR, T2, PD, and T1 sequences within new lesions, as well as two models to relate the clinical information to the four profiles.  The methodology presented here shows promise for both understanding the time course of tissue damage in the disease MS and may prove to be useful for evaluating the impact of neuroprotective or reparative treatments for the disease, for example in clinical trials.  Indeed, reliable methods to evaluate such treatments, which are currently in development, are lacking at present.   In contrast to prior studies of change in lesion intensity in clinical trials, our work is focused on voxel level analysis, and therefore it can provide spatial information about intensity recovery and does not artificially reduce the size of the data set. This may have implications on the sample size calculations for clinical trials. These methods are also broadly applicable to other imaging modalities and disease areas, in which longitudinal intensity profiles may lead to more sensitive and specific biomarkers. 

The inference from both of the models in regards to disease-modifying treatment should only be taken as a proof-of-concept for the relationship between the imaging and the clinical covariates.  The models may suffer from confounding by indication, which arises when individuals who are on a treatment are different from those who do not receive treatment, due to unobserved considerations.  In the multivariate model, we adjust for age, sex, and disease subtype, but unobservable differences related to treatment choice may cause biased results. However, bias in terms of treatment effect would most plausibly result in underestimation of improvements, as more aggressive therapies are usually given to subjects with more aggressive or refractory disease. Thus, our findings might underestimate what would be observed in a randomized trial of disease-modifying therapy. 

While many of the coefficient functions from the function-on-scalar regression are not found to be statistically different from 0, this model may have more power with more subjects.  For the bootstrap procedure we only have 34 subjects, resulting in wide confidence intervals for the estimated coefficient functions.  In contrast, the regression using score outcomes identifies strong associations between specific covariates and longitudinal patterns of longitudinal intensities.

The two models presented in this work are fit voxel-wise and therefore may be sensitive to major misregistration within a study and between longitudinal studies for the same subject.  
The models are also sensitive to local displacement of tissue due to transient swelling in and around or resorption of lesion tissue.  We therefore do not call the slow changes in intensity within the voxels that are observed ``tissue repair," as we cannot be certain that the change is not due to misregistration or displacement of tissue from the lesions themselves.  We do observe a relationship between the return of voxels to the intensity of normal-appearing tissue and both disease-modifying treatment and treatment with steroids, and therefore find this measure useful and deserving of further study.  We also see an association with the distance to the boundary of the lesion and slow, long-termintensity changes -- with voxels near the boundary of the lesion returning to baseline intensity and voxels near the center of the lesion maintaining abnormal signal intensity.   Future work to assess tissue repair may involve investigating a nonlinear registration within individual lesions.

The methods described here use only conventional clinical imaging for patients with MS, namely FLAIR, T2, PD, and T1.  While this is beneficial for using the methodology in a clinical trial setting or for analysis of retrospective imaging studies, one could also incorporate advanced imaging into the method.  For example, magnetization transfer ratio imaging \citep{van1999axonal}, quantitative T1-weighted imaging \citep{filippi2000quantitative}, and diffusion tensor imaging \citep{filippi2001diffusion} have been studied in MS lesions.  The longitudinal dynamics of lesions on these images could be incorporated into our framework to better understand the behavior of lesions over time and the impact of disease-modifying therapies on this behavior. 

For this analysis, all MRI studies are acquired on a single 1.5 T MRI scanner at one imaging center.  Similar analysis could be performed at higher field strength, but for this analysis we use a 1.5 T dataset for the availability of the large retrospective cohort study over a long period of time.  Although different scanning parameters were used for the acquisitions, further investigation is warranted into the robustness of the methods to changes in scanner, changes in magnetic field strength, as well changes in the imaging center.

\section*{Acknowledgments}  The project described is supported in part by the NIH grants RO1 EB012547 from the National Institute of Biomedical Imaging and Bioengineering,  RO1 NS060910 and RO1 NS085211  from the National Institute of Neurological Disorders and Stroke (NINDS),  T32 AG021334 from the National Institute of Aging, and RO1 MH095836 from the National Institute of Mental Health.  The research is also supported by the Intramural Research Program of NINDS.   The content is solely the responsibility of the authors and does not necessarily represent the official views of the funding agencies

\newpage 

 \bibliography{mybib}
\bibliographystyle{apalike}

\newpage

\section{Appendix} 

\subsection*{Parametric Bootstrapping Procedure} 

Let $B$ be the number of bootstrap samples to be performed and let $b$ index these $B$ samples.  Let $Y_{ilv}$ be the outcome for an observation indexed by $i$, $l$, and $v$.  Let $\boldsymbol{X}$ be the design matrix and $\boldsymbol{\beta}$ be the vector of the coefficients.  For this analysis we have a model of the form: 
\begin{align} 
Y_{ilv}  & = \boldsymbol{X \beta} + b_i + b_l + \epsilon_{ilv}  \nonumber 
\end{align} 

\noindent where $b_i \sim N \left(0, \sigma_{i}^2 \right)$ and $b_l \sim N \left(0, \sigma_{l}^2 \right)$ are random intercepts, and $\epsilon_{ilv} \sim N \left(0, \sigma_{\epsilon}^2  \right)$ is an error term.  For the parametric model, we fit the above mixed-effect model to get an estimate of $\boldsymbol{\beta}$, which we denote as $\boldsymbol{\hat \beta}$.  We then fix this estimate, and keep $\boldsymbol{X \hat \beta}$.  Using the fitted variances, $\hat \sigma_i^2$, $\hat \sigma_l^2$ and $\hat \sigma_{\epsilon}^2$,  we generate a random intercept for each lesion from a $N \left(0, \hat \sigma_{i}^2 \right)$ distribution, a random intercept for each subject from a $N \left(0, \hat \sigma_{l}^2 \right)$, and random noise for each voxel from a $N \left(0, \hat \sigma_{\epsilon}^2 \right)$.  We then add the random intercepts and noise to $\boldsymbol{X \hat \beta}$ for the corresponding observation and use this as our outcome to refit the model and get out bootstrapped coefficient vector  $\boldsymbol{\beta_{b} ^{*}}$.  To obtain the bootstrap sample, we repeat this procedure $B$ times.

\subsection*{Expert Validation} 

\begin{figure}[h!]
   \begin{center}
 \includegraphics[width=6in]{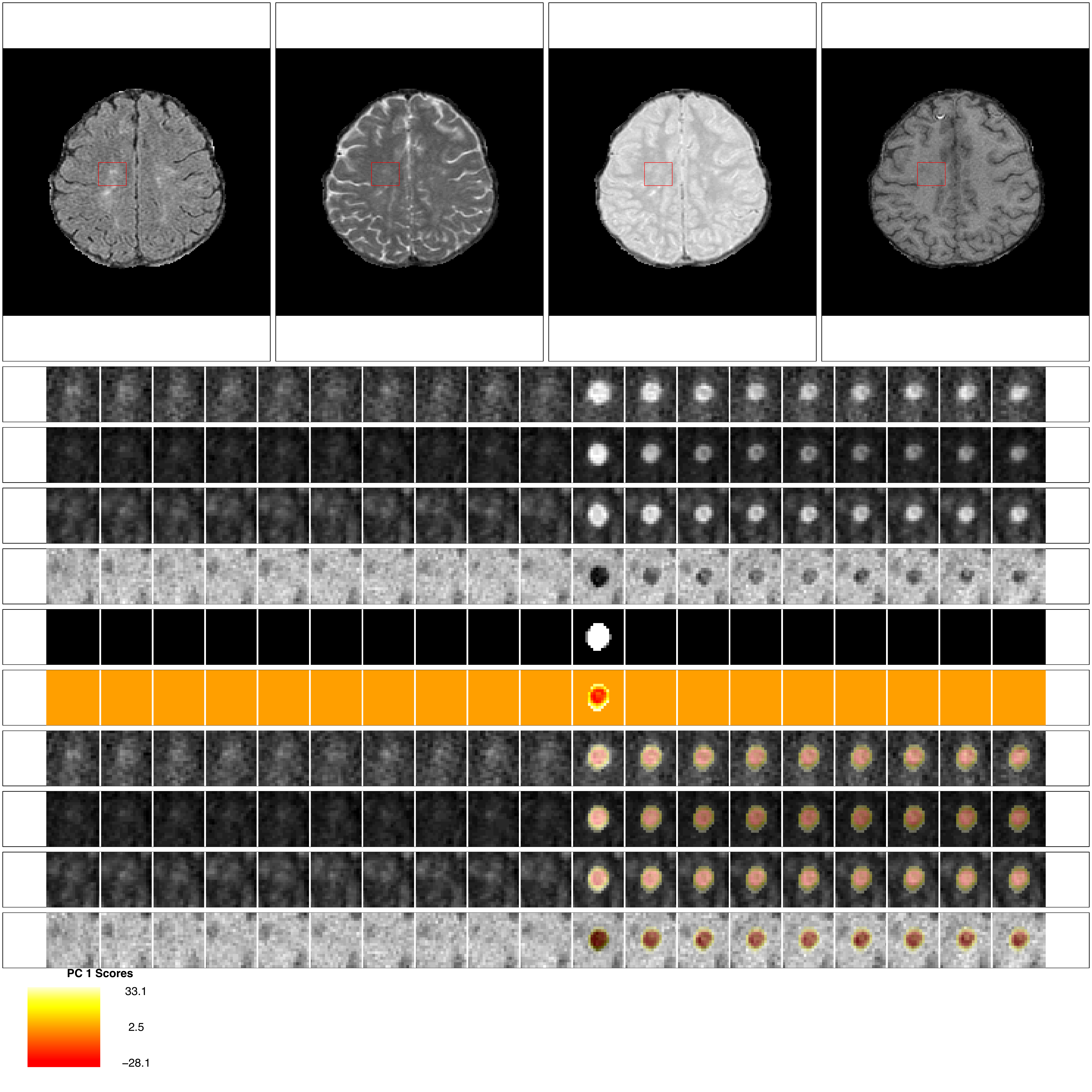}
 \end{center}
 \captionof{figure}{ \textbf{Passed: Rating of 4 for the score on the first PC. }  This scan received a rating of 4 for the score on the first PC from both raters.  Both raters also gave a rating of 4 for the lesion segmentation. }
       \label{fig:rate4}
   \end{figure}

\begin{figure}[h!]
   \begin{center}
 \includegraphics[width=6in]{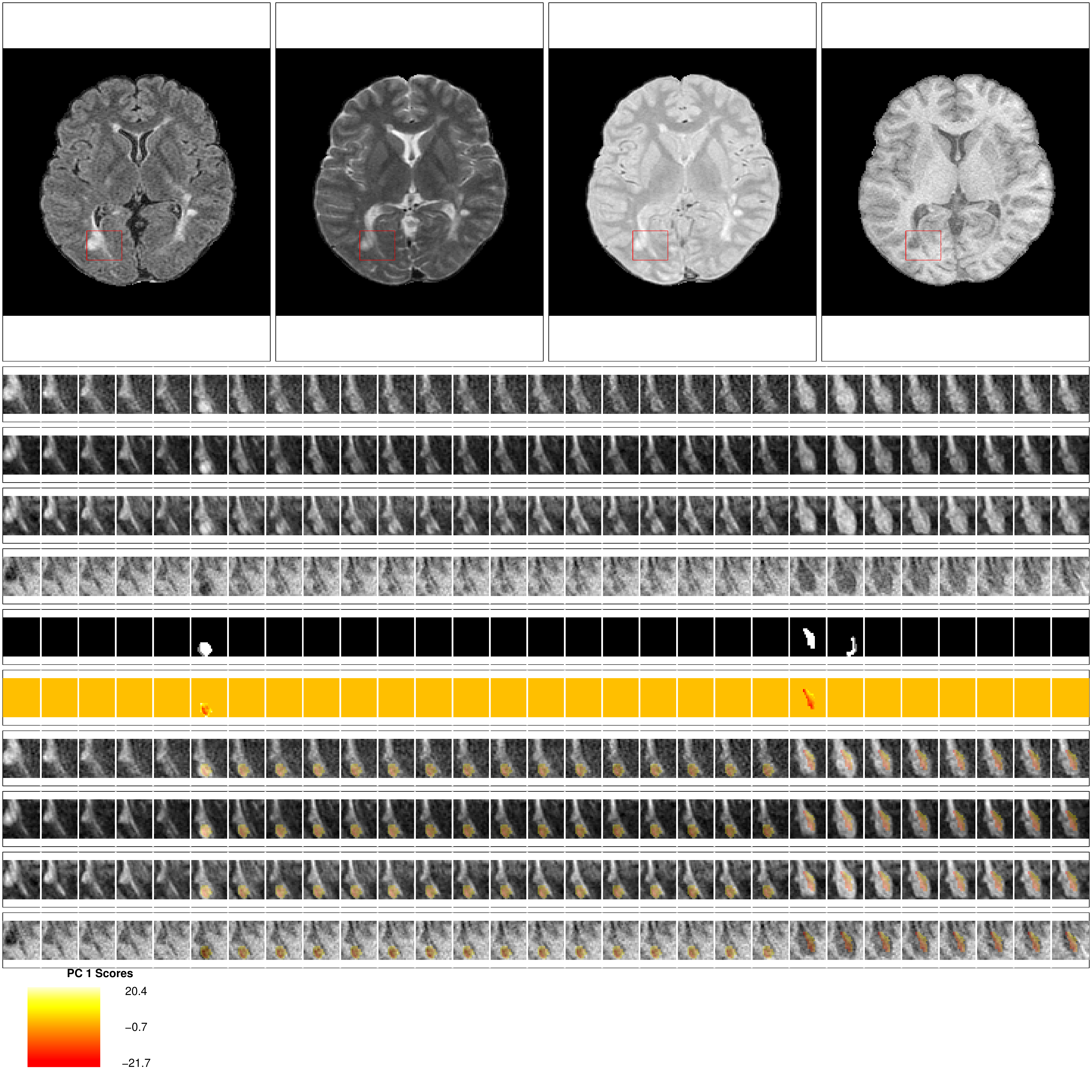}
 \end{center}
 \captionof{figure}{ \textbf{Passed with minor errors: Rating of 3 for the score on the first PC.} This scan received a rating of 3 for the score on the first PC from both raters.  Both raters also gave a rating of 3 for the lesion segmentation.  Note that at the 23rd time point new lesion voxels are segmented, but the score for the first PC is not produced for this time point, as the voxels did not meet the scanning criteria for being included in the analysis, which is described in detail in the Methods section of the paper. }
       \label{fig:rate3}
   \end{figure}

\begin{figure}[h!]
   \begin{center}
 \includegraphics[width=6in]{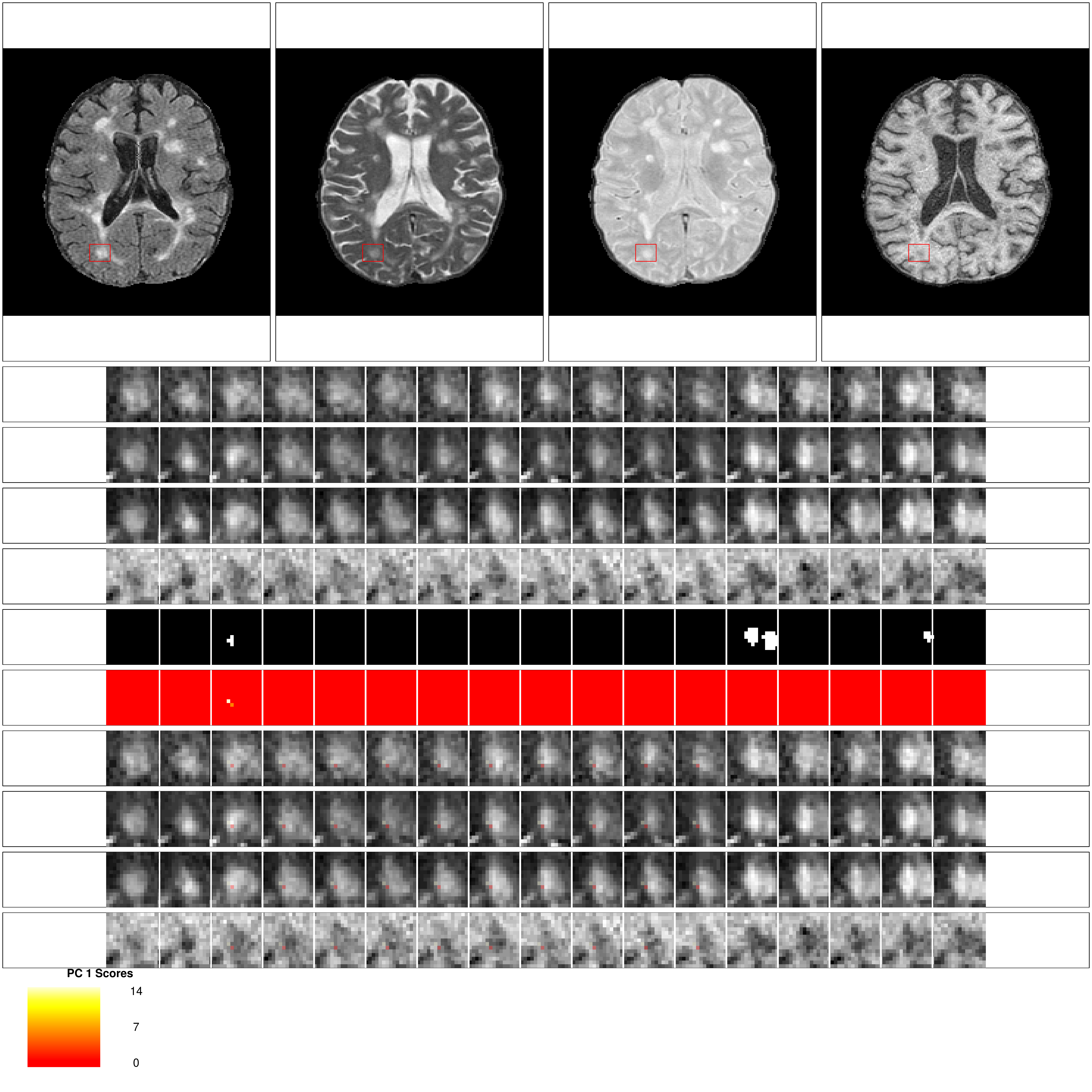}
 \end{center}
 \captionof{figure}{  \textbf{Some redeeming features: Rating of 2 for the score on the first PC.}  This scan received a rating of 2 for the score on the first PC from both raters.  Both raters also gave a rating of 2 for the lesion segmentation.  The neuroradioloigst commented that this scan received a low rating because it was not clear that the segmented portion for time point 3 was lesion.}
       \label{fig:rate2}
   \end{figure}

\begin{figure}[h!]
   \begin{center}
 \includegraphics[width=6in]{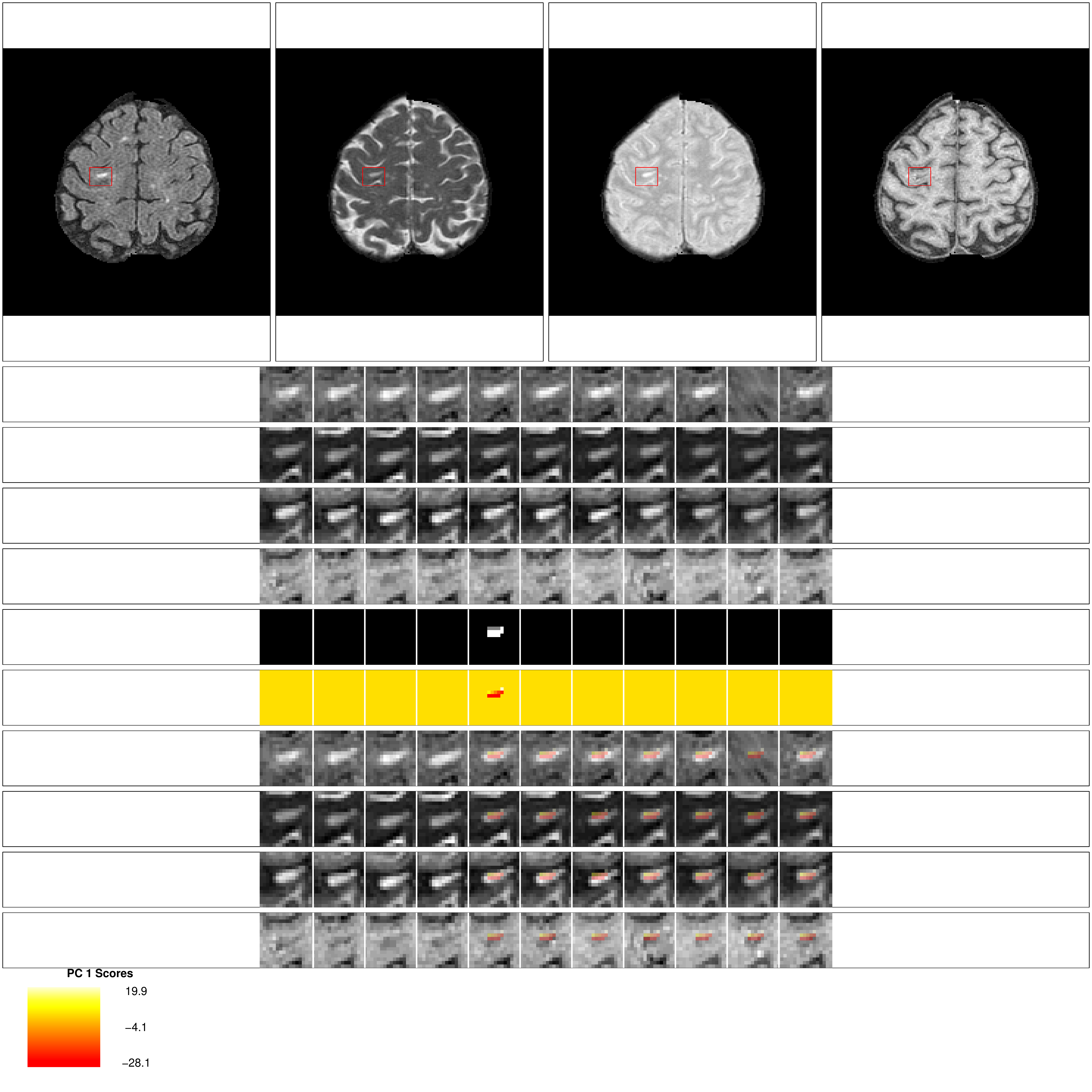}
 \end{center}
 \captionof{figure}{  \textbf{Failed miserably: Rating of 1 for the score on the first PC.} This scan received a rating of 1 for the score on the first PC from both raters.  Both raters also gave a rating of 1 for the lesion segmentation.  Both raters commented that the low rating was because the lesion had existed in all time points and was not a new lesion.  }
       \label{fig:rate1}
   \end{figure}

Examples of the set of evaluation images presented to the experts for each lesion are shown in Figures~\ref{fig:rate4}, \ref{fig:rate3},  \ref{fig:rate2}, and \ref{fig:rate1}. The first row of the figures shows the full axial slice for the FLAIR,  T2, PD, and T1 volumes that contains the largest number of voxels with abnormal signal intensity.  The second through fourth rows show the entire collection of longitudinal scans for a box containing the abnormal signal intensity in the FLAIR, T2, PD, and T1 weighted volumes for this axial slice.  The scans are displayed in chronological order, from first time point to last time point, from left to right.  The fifth row shows the segmentation of the lesion and edema tissue within this box at each time point. The sixth row shows the score on the first PC for the voxels segmented as lesion tissue, displayed at the time of lesion incidence for each voxel.  The seventh throughout tenth row shows  the entire collection of longitudinal scans for the FLAIR, T2, PD, and T1 weighted volumes within this box with the score for the first PC overlaid on the images for each scan after lesion incidence.  The last row shows the scale for the score on the first PC.  The figures show examples of the four different ratings for the score on the first PC.  Both raters rate the scans as either (1) failed miserably, (2) some redeeming features, (3) passed with minor errors or (4) passed.

\bigskip 

   \newpage    
  
  \newpage     
  
  \newpage    
    
 \newpage         
   
\subsection*{Principal Component Analysis and Regression} 

Table~\ref{table:multivariate} shows the coefficient estimates, standard errors, t-statistics, the p-values using the normal approximation, and the 95\% bootstrapped confidence intervals for the multivariate PCA regression model.  Table~\ref{table:univariate} shows the same for the individual univariate PCA regression models. 

\begin{table}[ht]
\centering
\begin{tabular}{rrrrrr}
  \hline
 & Estimate & Standard Error & t-value & p-value  & 95\% Bootstrapped CI \\ 
  \hline
SPMS & 2.15 & 4.41 & 0.49 & 0.63 & (-6.19, 10.93)  \\ 
  Distance to Boundary  & -9.39 & 0.08 & -123.74 & 0.00 & (-9.56, -9.25)  \\ 
 Age    & -0.21 & 0.18 & -1.16 & 0.25 & (-0.57,  0.13) \\ 
 (Age $-$ 4)$_{+}$     & -0.10 & 0.23 & -0.42 & 0.68 & (-0.54, 0.35)  \\ 
   Steroids & 4.26 & 0.79 & 5.42 & 0.00 & (2.67, 5.85) \\ 
 Male   & 1.16 & 2.55 & 0.45 & 0.65  & (-3.94, 6.61) \\ 
Treatment  & 5.39 & 0.36 & 15.03 & 0.00   & (4.67, 6.08) \\ 
Intercept & 8.89 & 1.92 & 4.64 & 0.00 & (5.17, 12.85) \\ 
   \hline
\end{tabular}
\caption{\textbf{Coefficient estimates, standard errors, t-statistics, p-values, and bootstrapped 95\% confidence intervals for the multivariate PCA regression model.}}
       \label{table:multivariate}
\end{table}

\begin{table}[ht]
\centering
\begin{tabular}{rrrrrr}
  \hline
 & Estimate & Standard Error & t-value & p-value & 95\% Bootstrapped CI \\ 
  \hline
SPMS & 0.65 & 4.11 & 0.16 & 0.88 &  ( -7.71, 9.18)  \\ 
  Distance to Boundary  & -9.37 & 0.08 & -123.18 & 0.00  &(-9.52, -9.22) \\ 
 Age   & 0.89 & 0.19 & 4.58 & 0.00 & (0.51, 1.23)  \\ 
 (Age $-$ 4)$_{+}$    & -1.55 & 0.24 & -6.40 & 0.00 & (-1.95, -1.14) \\ 
   Steroids  & 6.03 & 0.78 & 7.77 & 0.00  & (4.55, 7.59) \\ 
 Male  & 0.43 & 2.43 & 0.18 & 0.86 & (-4.32, 4.97) \\ 
Treatment & 4.48 & 0.38 & 11.76 & 0.00 & (3.67, 5.25)  \\ 
   \hline
\end{tabular}
\caption{\textbf{Coefficient estimates, standard errors, t-statistics, p-values, and bootstrapped 95\% confidence intervals for the univariate PCA regression model.}}
       \label{table:univariate}
\end{table}

\subsection*{Density Plots} 

Density plots for the score on the first PC by disease subtype, treatment with steroids and disease modifying therapies, and sex are shown below.  A plot of the score on the first PC versus age is also shown below.  Disease subtype and sex were not found to be statistically significant in the linear mixed-effects model, as there were few subjects in our analysis with SPMS or that were male. 

\begin{figure}[h!]
   \begin{center}
 \includegraphics[width=5in]{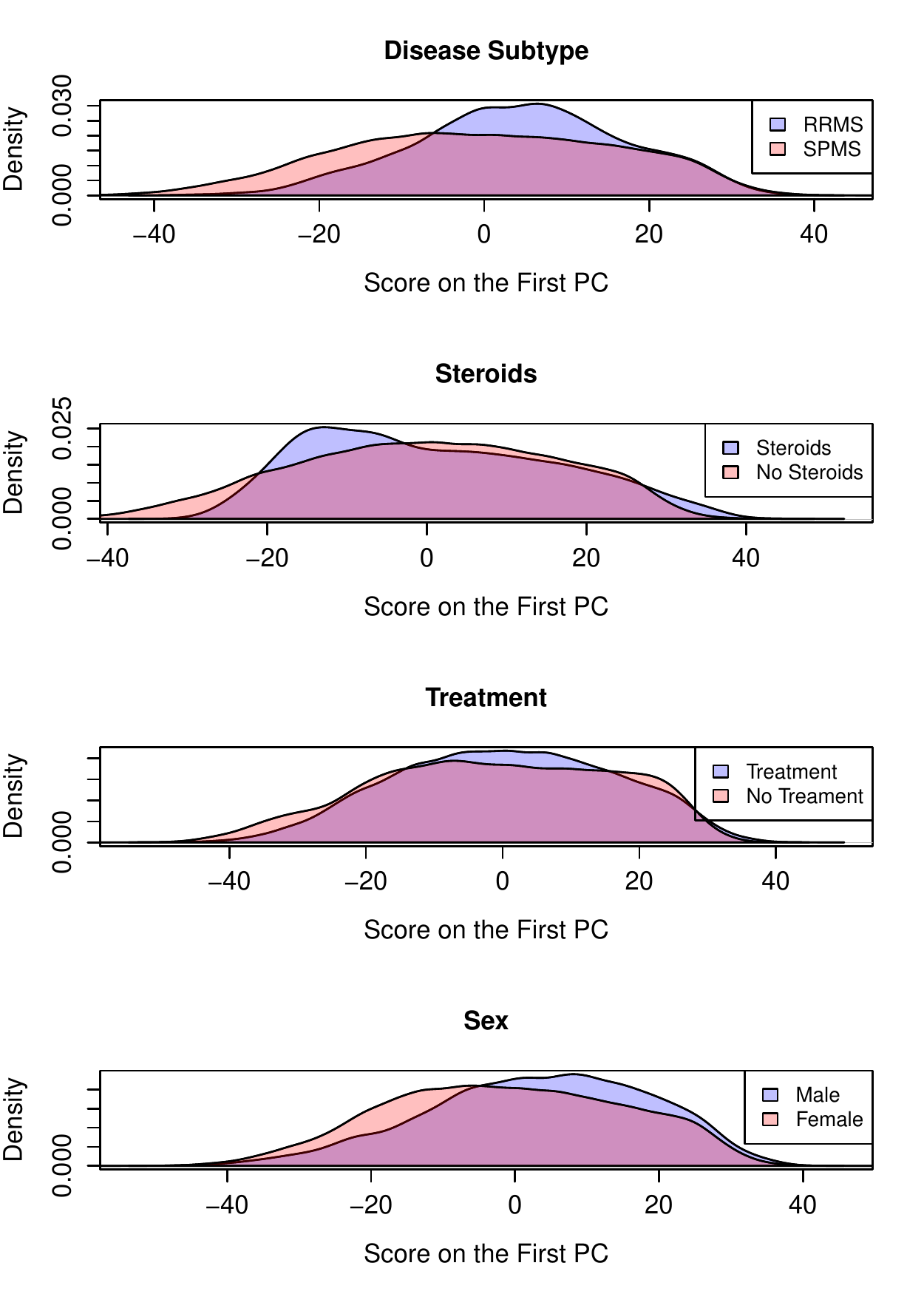}
 \end{center}
 \captionof{figure}{ \textbf{Density plots for the score on the first PC.}  Each plot shows the density for the score on the first PC.  In the first row the density is shown by disease subtype, the second row by treatment with steroids, the third row by disease-modifying treatment, and the fourth row by sex.    }
       \label{fig:den}
   \end{figure}

\begin{figure}[h!]
   \begin{center}
 \includegraphics[width=5in]{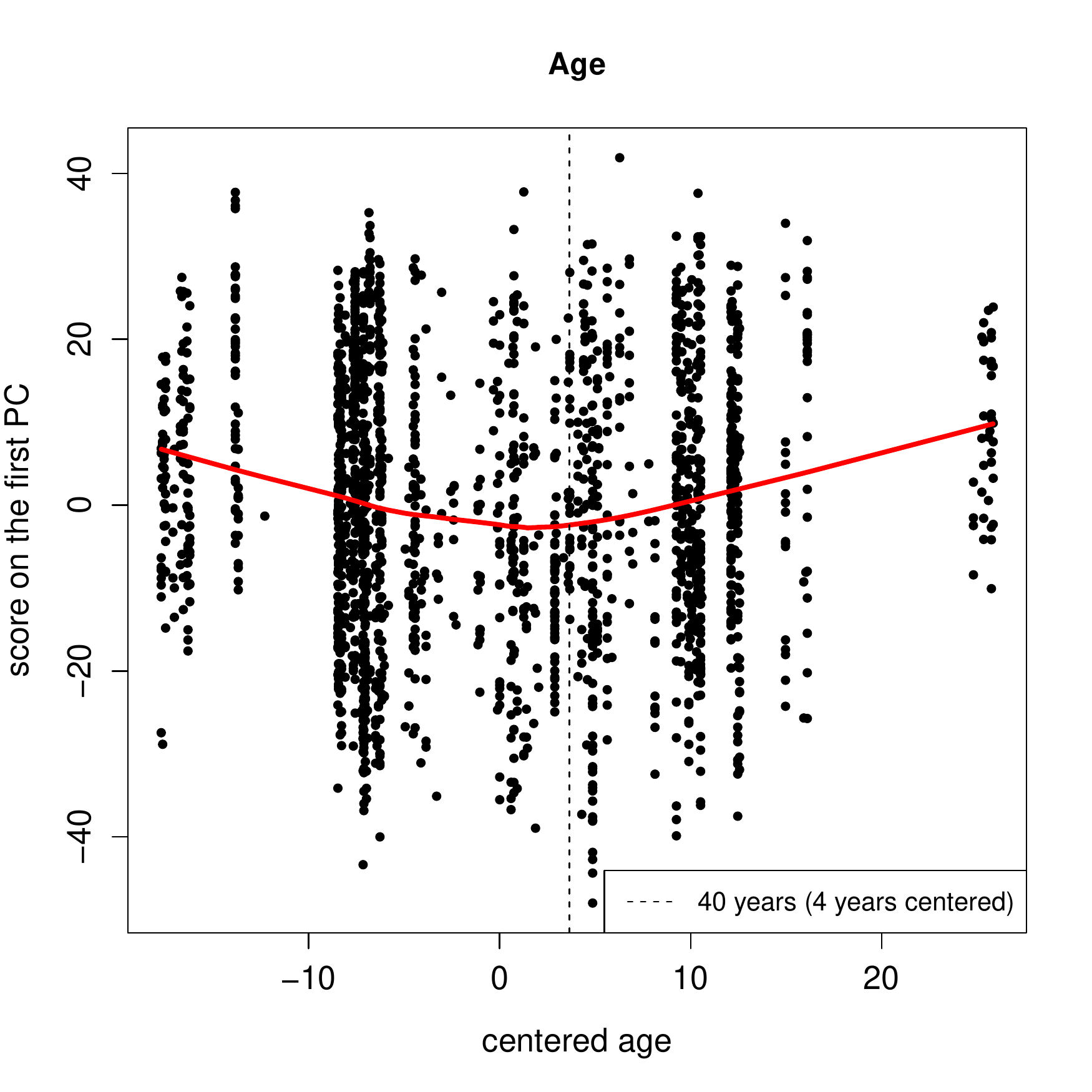}
 \end{center}
 \captionof{figure}{ \textbf{Score on the first PC versus age.}  The score on the first PC versus the centered age of the subject (at the time time of lesion incidence) for each voxel.  For purposes of presentation 2000 voxels were sampled at random for the plot.  The red line shows the locally weighted scatterplot smoothing through the entire dataset. The dashed line shows the where the spline term for age was added at 40 years of age (or 4 years of age for centered age). }
       \label{fig:fosrt2}
   \end{figure}

  \newpage 

\subsection*{Function-on-Scalar Regression} 

The coefficient functions from the function-on-scalar regression with bootstrapped $95\%$ confidence intervals with the T2, PD, and  T1 profile as the outcome are shown below. Similar to using the FLAIR profile as the outcome, only the distance to the boundary and age were found to be different from 0 at any point along the profile. 

\begin{figure}[h!]
   \begin{center}
 \includegraphics[width=5in]{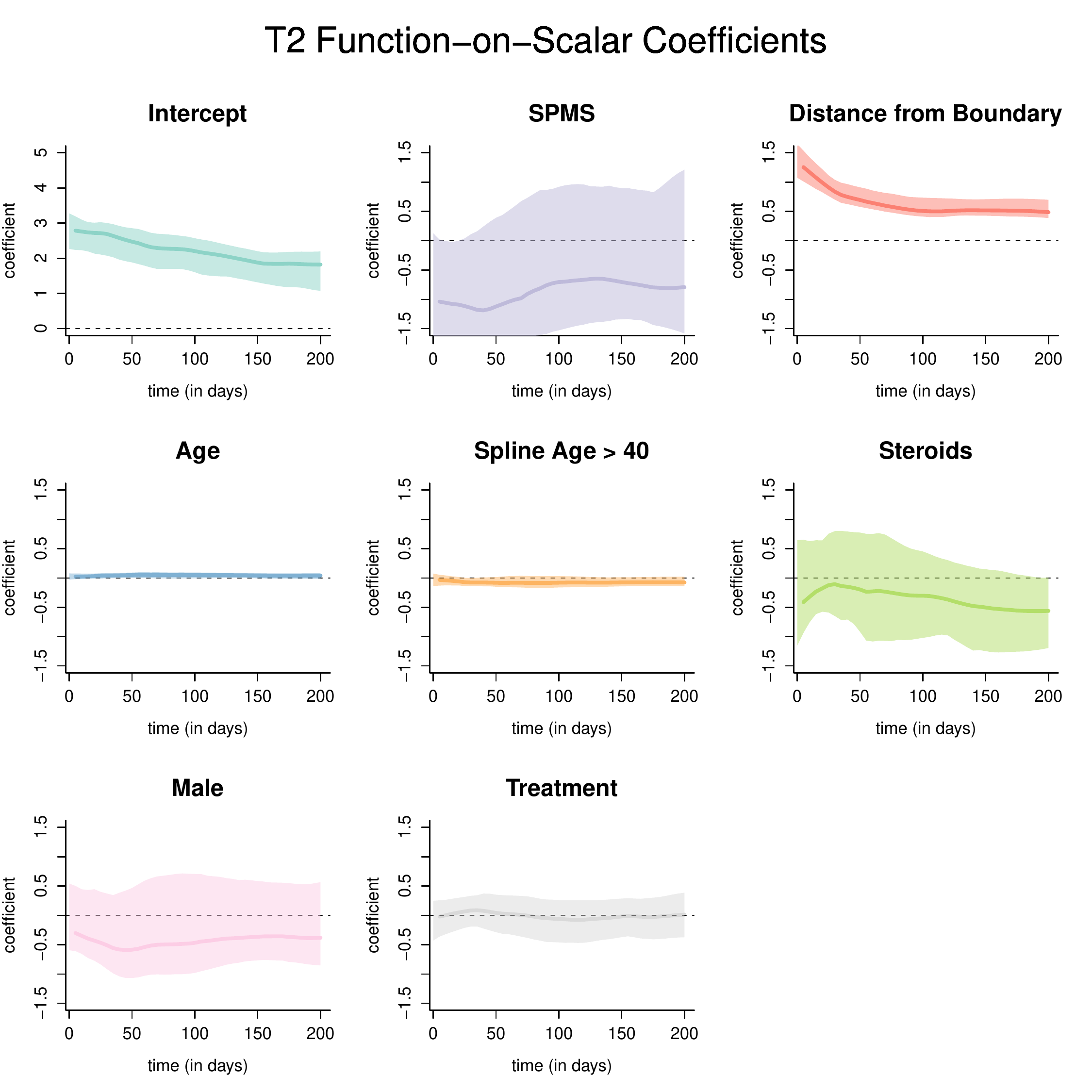}
 \end{center}
 \captionof{figure}{ \textbf{Coefficient functions from the function-on-scalar regression with the T2 profile as an outcome.}   Each dark line represents the coefficient function, and the shaded area represents a bootstrapped, point wise 95\% confidence interval.  Along the y-axis is the value of the coefficient function at each time point. Only distance from the boundary and age were found to be different from 0 at any point along the profile. }
       \label{fig:fosrt2}
   \end{figure}

   \begin{figure}[h!]
   \begin{center}
 \includegraphics[width=5in]{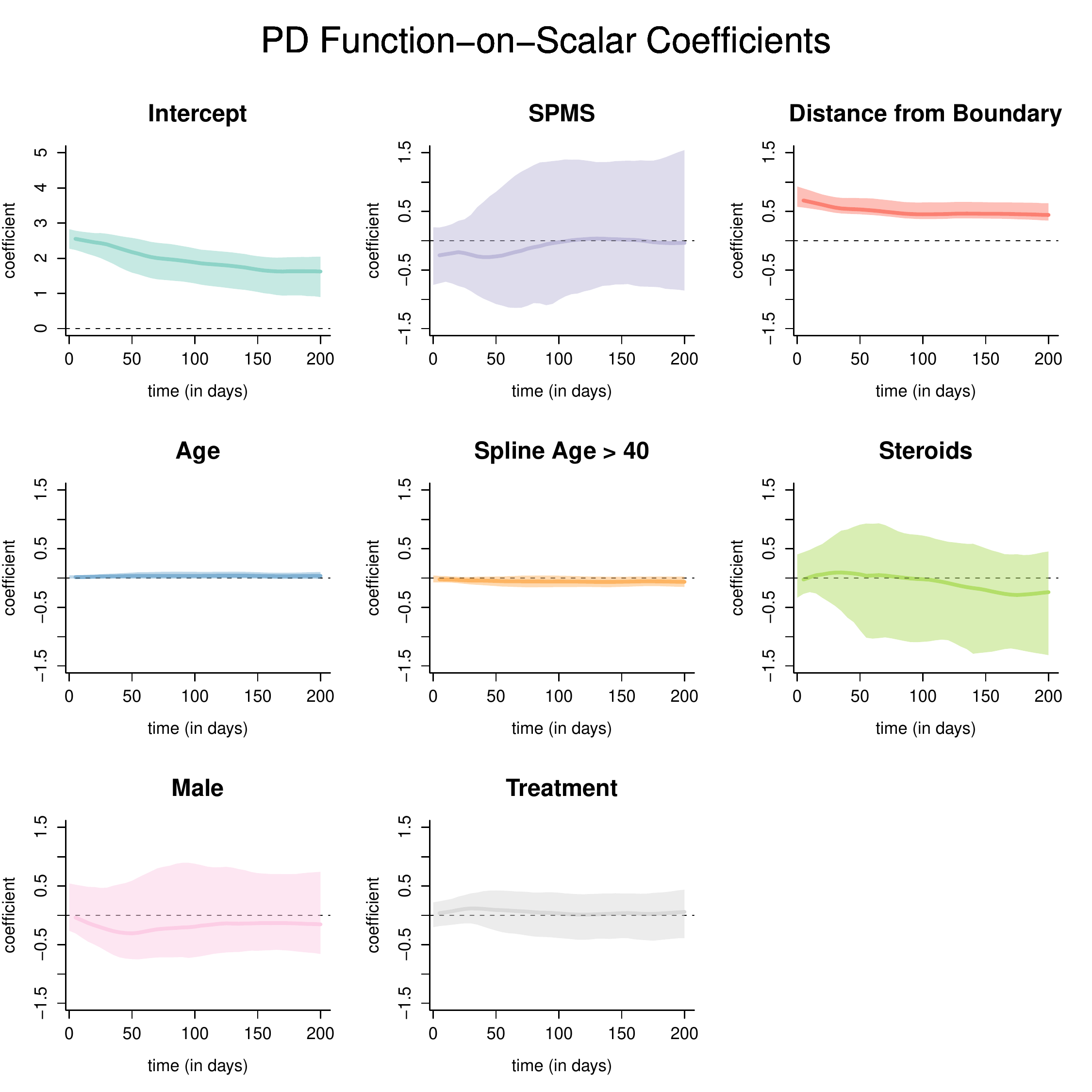}
 \end{center}
 \captionof{figure}{ \textbf{Coefficient functions from the function-on-scalar regression with the PD profile as an outcome.}   Each dark line represents the coefficient function, and the shaded area represents a bootstrapped, point wise 95\% confidence interval.  Along the x-axis of each plot is the time in days from lesion incidence.  Along the y-axis is the value of the coefficient function at each time point. Only distance from the boundary and age were found to be different from 0 at any point along the profile. }
       \label{fig:fosrpd}
   \end{figure}

   \begin{figure}[h!]
   \begin{center}
 \includegraphics[width=5in]{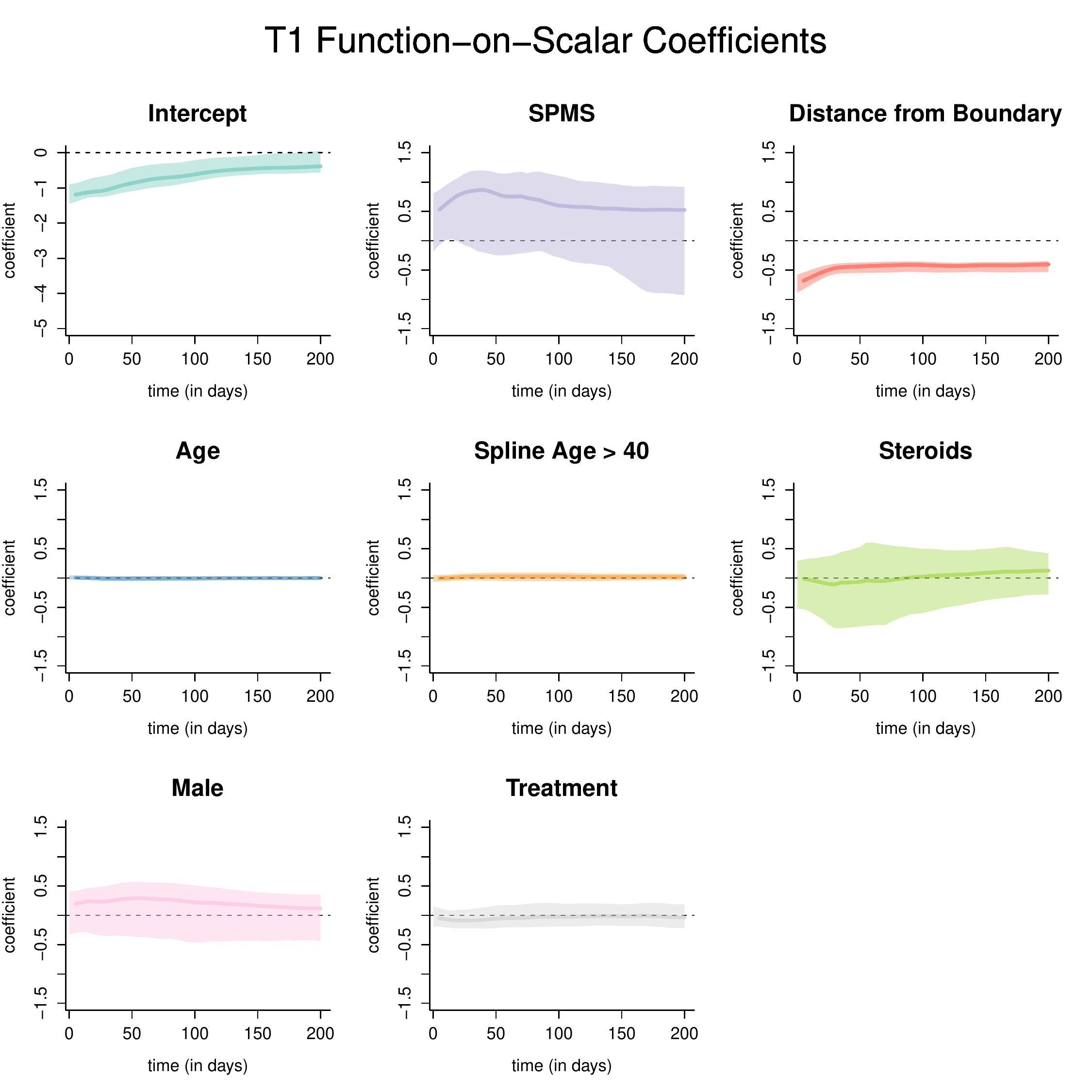}
 \end{center}
 \captionof{figure}{ \textbf{Coefficient functions from the function-on-scalar regression with the T1 profile as an outcome.}  E Each dark line represents the coefficient function, and the shaded area represents a bootstrapped, point wise 95\% confidence interval.  Along the x-axis of each plot is the time in days from lesion incidence.  Along the y-axis is the value of the coefficient function at each time point. Only distance from the boundary and age were found to be different from 0 at any point along the profile. }
       \label{fig:fosrt1}
   \end{figure}

\end{document}